\documentclass[%
 reprint,
superscriptaddress,
 amsmath,amssymb,
 aps,
]{revtex4-2}

\usepackage{graphicx}
\usepackage{dcolumn}
\usepackage{bm}
\usepackage{color}
\usepackage[colorlinks,linkcolor=blue,citecolor=blue]{hyperref}
\usepackage{bm}
\usepackage{amsmath}
\usepackage{float}

\begin{document}

\preprint{APS/123-QED}

\title{Topologically protected subradiant cavity polaritons through linewidth narrowing enabled by dissipationless edge states}

\author{Yuwei Lu}
\email[Corresponding Author: ]{luyw5@fosu.edu.cn}
\affiliation{School of Physics and Optoelectronics, South China University of Technology, Guangzhou 510641, China}
\affiliation{School of Physics and Optoelectronic Engineering, Foshan University, Foshan 528000, China}

\author{Jingfeng Liu}%
\affiliation{College of Electronic Engineering, South China Agricultural University, Guangzhou 510642, China}

\author{Haoxiang Jiang}%
\affiliation{Guangdong Provincial Key Laboratory of Quantum Engineering and Quantum Materials, School of Physics and Telecommunication Engineering, South China Normal University, Guangzhou 510006, China}
\affiliation{Guangdong-Hong Kong Joint Laboratory of Quantum Matter, Frontier Research Institute for Physics, South China Normal University, Guangzhou 510006, China}


\author{Zeyang Liao}
\email[Corresponding Author: ]{liaozy7@mail.sysu.edu.cn}
\affiliation{State Key Laboratory of Optoelectronic Materials and Technologies, School of Physics, Sun Yat-sen University, Guangzhou 510275, China}



\begin{abstract}
Cavity polaritons derived from the strong light-matter interaction at the quantum level provide a basis for efficient manipulation of quantum states via cavity field. Polaritons with narrow linewidth and long lifetime are appealing in applications such as quantum sensing and storage. Here, we propose a prototypical arrangement to implement a whispering-gallery-mode resonator with topological mirror moulded by one-dimensional atom array, which allows to boost the lifetime of cavity polaritons over an order of magnitude. This considerable enhancement attributes to the coupling of polaritonic states to dissipationless edge states protected by the topological bandgap of atom array that suppresses the leakage of cavity modes. When exceeding the width of Rabi splitting, topological bandgap can further reduce the dissipation from polaritonic states to bulk states of atom array, giving arise to subradiant cavity polaritons with extremely sharp linewidth. The resultant Rabi oscillation decays with a rate even below the free-space decay of a single quantum emitter. Inheriting from the topologically protected properties of edge states, the subradiance of cavity polaritons can be preserved in the disordered atom mirror with moderate perturbations involving the atomic frequency, interaction strengths and location. Our work opens up a new paradigm of topology-engineered quantum states with robust quantum coherence for future applications in quantum computing and network. 
\end{abstract}

\maketitle


\section{Introduction}

Cavity quantum electrodynamics (QED) constitutes one of the cornerstones of quantum optics, where the coherent exchange of single photon between the quantum emitter (QE) and the cavity mode, known as Rabi oscillation, can take place in the strong-coupling regime and results in the formation of polaritonic states consisting of entangled atom and photon components \cite{RN2,RN3}. The corresponding bosonic quasiparticle, termed cavity polaritons, offers a scheme for controllable storage and transfer of quantum states and a rich variety of technologies and applications, such as on-chip quantum light source \cite{RN4,RN6}, quantum sensing \cite{RN7,RN8}, scalable quantum computing and quantum information processing \cite{RN3,RN10,RN11,RN12}. Great effort has been devoted into achieving strong coupling in various QED platforms \cite{RN13,RN14,RN16,RN17}, while less attention has been paid to reduce the linewidth of cavity polaritons \cite{RN19,RN20,RN22,PRRFano}, which is beneficial for diverse quantum-optics applications \cite{RN23,RN24,RN25,RN26,RN27,RN28,RN29}. For instance, reducing the linewidth of resonant systems enables to detect weak signals and achieves better measurement sensitivity for precision sensing in experiments \cite{RN27,RN28,RN29,RN30,RN32,LOE}. Moreover, linewidth represents decay rate, thereby quantum states with narrower linewidth means longer lifetime, a feature highly desirable for quantum storage and quantum memory \cite{RN33,RN34,RN35,RN36,RN37}. The lifetime of cavity polaritons is often limited by the quality ($Q$) factor of cavity, since the linewidth of QE is usually smaller than that of cavity in many cavity-QED systems \cite{RN14,RN17,RN38,RN39}. However, a high-$Q$ cavity in general features a large volume \cite{RN40,RN42,RN43} or requires sophisticated design \cite{RN45,RN46} that is demanding for nanofabrication.
\begin{figure*}[t]
\centering\includegraphics[width=0.76\linewidth]{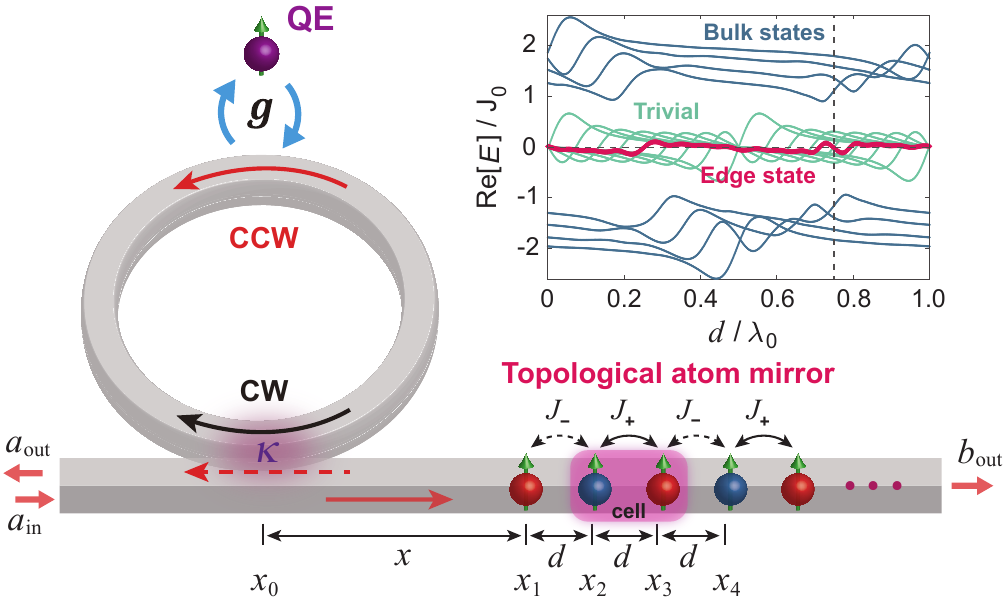}
\caption{Schematic of a whispering-gallery-mode (WGM) ring cavity coupled to a quantum emitter (QE) and a waveguide with a one-dimensional topological atom mirror at the right end. $x_j$ indicates the location of the $j$th element. The staggered hoppings between nearest-neighbor sites in topological atom mirror simulate the Su-Schrieffer-Heeger (SSH) chain that supports the topological edge states. The pair of sites with stronger coupling defines a unit cell, as the pink translucent box indicates ($J_{+}>J_{-}$). The inset shows the real energy spectrum of topological atom mirror versus atom spacing $d$ for nine atoms. Vertical dashed line indicates the dissipationless topological edge state at $d=3\lambda_0/4$ under investigation. Other parameters are $J_0=8\Gamma$, $\phi_1 = 0$ and $\phi=0.3\pi$. $a_{in}$ and $a_{out}$, $b_{out}$ stand for the input and output fields for planewave excitation, respectively. }
\label{fig1}
\end{figure*}

Beyond the conventional quantum optics, recently topological quantum optics appeared as a rapidly growing field for controlling light-matter interaction in many-body quantum systems by exploiting the concept of topology \cite{RN47,RN48,RN49,RN50,RN51,RN52,RN53,RN54,RN55,RN56,RN57}. In analogy to photonic topological insulators, the emergence of exotic topological states in quantum systems, characterized by localized edge states and interface states, demonstrates intriguing optical response and has motivated the development of functional quantum devices with robustness against the structural disorder and impurities, such as topological single-photon circulator \cite{RN59}, topologically protected qubits \cite{RN49,RN52,RN63}, unconventional photon transport \cite{RN54,RN56,RN64}, fault-tolerant topological quantum computing \cite{RN65,RN66,PRXQuantum}, to mention a few. Among these topological quantum systems, atom arrays can serve as a versatile platform for topological light manipulation, with functionalities beyond the classical mirror that reflects the light \cite{RN49,RN51,PRResearch.2.033190,PRA.105.043514}. In particular, topological quantum states can become subradiant through the collective interference \cite{RN49,RN51,RN63}, whose radiative loss can be strongly suppressed and significantly smaller than the free-space decay rate of a single atom. This unique feature of atom arrays combined with topological protection provides extra degrees of freedom to manipulate the quantum states.  

Triggered by the prospect of manipulating cavity polaritons through topological effects, we propose a topological edge states-engineered cavity QED system consisting of a whispering-gallery-mode (WGM) resonator coupled to a one-dimensional (1D) topological atom mirror with long-range hoppings mediated by waveguide. With sufficiently strong atom-waveguide coupling, edge states become dissipationless through topological phase transition \cite{RN53}. By virtue of the exponential localization and topological protection of dissipationless edge states, this simple configuration enables unprecedented linewidth narrowing and decay suppression of polaritonic states with small atom array. By analyzing the energy spectrum and spectral properties of the composed system, we predict that typically a dozen atoms are adequate to produce subradiance for cavity polaritons and the resultant subnatural linewidth can be experimentally evidenced from either the reflection spectrum of waveguide or the fluorescence of QE. Our scheme can provide a viable approach to realize the long-time storage of quantum states in QED systems with cavities of moderate $Q$ factor and explore the topological manipulation of quantum states on integrated optoelectronic platform.

\section{Results and Discussion}
\subsection{Model and Theory}
The system under investigation is depicted in Fig. \ref{fig1} and comprises of a hybrid cavity QED system based on a WGM ring resonator and a waveguide QED system \cite{RN69}. The resonator supports a series of WGM resonances, but only a pair of degenerate clockwise (CW) and counterclockwise (CCW) modes with the same WGM order is considered. This simplification is reasonable for a realistic WGM resonator operated in the visible and near-infrared ranges, where the linewidth of QE can be much smaller than the frequency spacing between adjacent WGM resonances \cite{nl2007,prb2009,RN38,RN25}. A QE is embedded inside the resonator, which couples to a waveguide with a topological atom mirror at the right end. The nearest-neighbor interactions between topological atoms change alternatively to form 1D diatomic chain by mimicking the Su-Schrieffer-Heeger (SSH) model \cite{RN47,RN68}, in addition to the long-range interactions mediated by waveguide. An extended cascaded quantum master equation is derived in Appendix \ref{aa} and employed to describe the quantum dynamics of the composed system (see also Refs. \cite{RN24,RN69} for details)
\begin{equation}\label{eq1}
\dot{\rho}=-i[H, \rho]+\mathcal{D}[\rho]
\end{equation}
with Lindblad operator
\begin{widetext}
    \begin{equation}\label{eq2}
    \begin{aligned}
    \mathcal{D}[\rho]=\frac{\kappa_R}{2} & \mathcal{L}\left[c_{c c w}\right] \rho+\frac{\kappa_L}{2} \mathcal{L}\left[c_{c w}\right] \rho+\sum_{j=0}^N \frac{\gamma_0}{2} \mathcal{L}\left[\sigma_{-}^{(j)}\right] \rho+\sum_{\lambda=R, L} \sum_{j=1}^N \frac{\gamma_\lambda}{2} \mathcal{L}\left[\sigma_{-}^{(j)}\right] \rho \\
    & + \sum_{\lambda=R, L} \sum_{\substack{j, l \geq 1 \\ k_\lambda x_j>k_\lambda x_l}}^N \gamma_\lambda\left(e^{i k_\lambda\left(x_j-x_l\right)}\left[\sigma_{-}^{(l)} \rho, \sigma_{+}^{(j)}\right]+e^{-i k_\lambda\left(x_j-x_l\right)}\left[\sigma_{-}^{(j)}, \rho \sigma_{+}^{(l)}\right]\right) \\
    & + \sum_{j=1}^N \sqrt{\kappa_R \gamma_R}\left(e^{i k_R x_j}\left[c_{c c w} \rho, \sigma_{+}^{(j)}\right]+e^{-i k_R x_j}\left[\sigma_{-}^{(j)}, \rho c_{c c w}^{\dagger}\right]\right) \\
    & + \sum_{j=1}^N \sqrt{\kappa_L \gamma_L}\left(e^{-i k_L x_j}\left[c_{c w} \rho, \sigma_{+}^{(j)}\right]+e^{i k_L x_j}\left[\sigma_{-}^{(j)}, \rho c_{c w}^{\dagger}\right]\right) \\
    &
    \end{aligned}
    \end{equation}
\end{widetext}

\noindent where the first line introduces the dissipation for individuals, the second line describes the waveguide-mediated interaction between atoms, and the third (four) line accounts for the chiral coupling between the atoms and the CCW (CW) mode through the right-propagating (left-propagating) guided mode of waveguide. $\mathcal{L}[O] \rho=2 O \rho O^{\dagger}-O^{\dagger} O \rho-\rho O^{\dagger} O$ is the Liouvillian superoperator for the dissipation of operator $O$. $c_{ccw}$ ($c_{cw}$) is the bosonic annihilation operator of CCW (CW) mode, while $\kappa_R$ ($\kappa_L$) is the corresponding decay rate stemming from the evanescent coupling to the waveguide. The intrinsic decay of cavity modes is omitted in consideration of the high-$Q$ feature of WGM resonators. $k_R=-k_L=k_0$ is the wave vector of photons. $\sigma_{-}^{(j)}$ is the lowering operator of the $j$th atom located at $x_j$ and particularly, $\sigma_{-}^{(0)}$ represents the atom inside the cavity, which we refer to QE hereafter to distinguish from the atoms in the mirror. $N$ and $x_0$ denote the number of atoms and the location of waveguide-cavity junction, respectively. $\gamma_0$ and $\gamma_\lambda$ ($\lambda=R,L$) stand for the free-space decay and the waveguide-induced decay of atoms, respectively. Throughout the paper, we consider symmetric coupling of atoms ($\gamma_R=\gamma_L=\Gamma$) and cavity modes ($\kappa_R=\kappa_L=\kappa$) to two chiral guided modes of waveguide. Meanwhile, the coherent interaction between atoms can be tailored by adjusting the atom spacing \cite{PRA2017,PRA2022}. Without loss of generality, we focus on the case of equal atom spacing, i.e., $x_{j+1}-x_j=d$ for $j \geq 1$. The total Hamiltonian reads
\begin{equation}\label{eq3}
H=H_0+H_I+H_{\text {topo }}
\end{equation}
with the free Hamiltonian
\begin{equation}
H_0=\omega_c c_{c c w}^{\dagger} c_{c c w}+\omega_c c_{c w}^{\dagger} c_{c w}+\omega_c \sigma_{+}^{(0)} \sigma_{-}^{(0)}+\sum_{j=1}^N \omega_j \sigma_{+}^{(j)} \sigma_{-}^{(j)}
\end{equation}
and the interaction Hamiltonian for cavity QED system
\begin{equation}
H_I=g\left(c_{c c w}^{\dagger} \sigma_{-}^{(0)}+\sigma_{+}^{(0)} c_{c c w}\right)+g\left(c_{c w}^{\dagger} \sigma_{-}^{(0)}+\sigma_{+}^{(0)} c_{c w}\right)
\end{equation}
and the Hamiltonian describing the coherent coupling between adjacent atoms
\begin{equation}\label{eq6}
H_{\mathrm{topo}}=\sum_{j=1}^{N-1} J_j\left(\sigma_{+}^{(j)} \sigma_{-}^{(j+1)}+\sigma_{+}^{(j+1)} \sigma_{-}^{(j)}\right)
\end{equation}

\begin{figure*}[t]
\centering\includegraphics[width=0.62\linewidth]{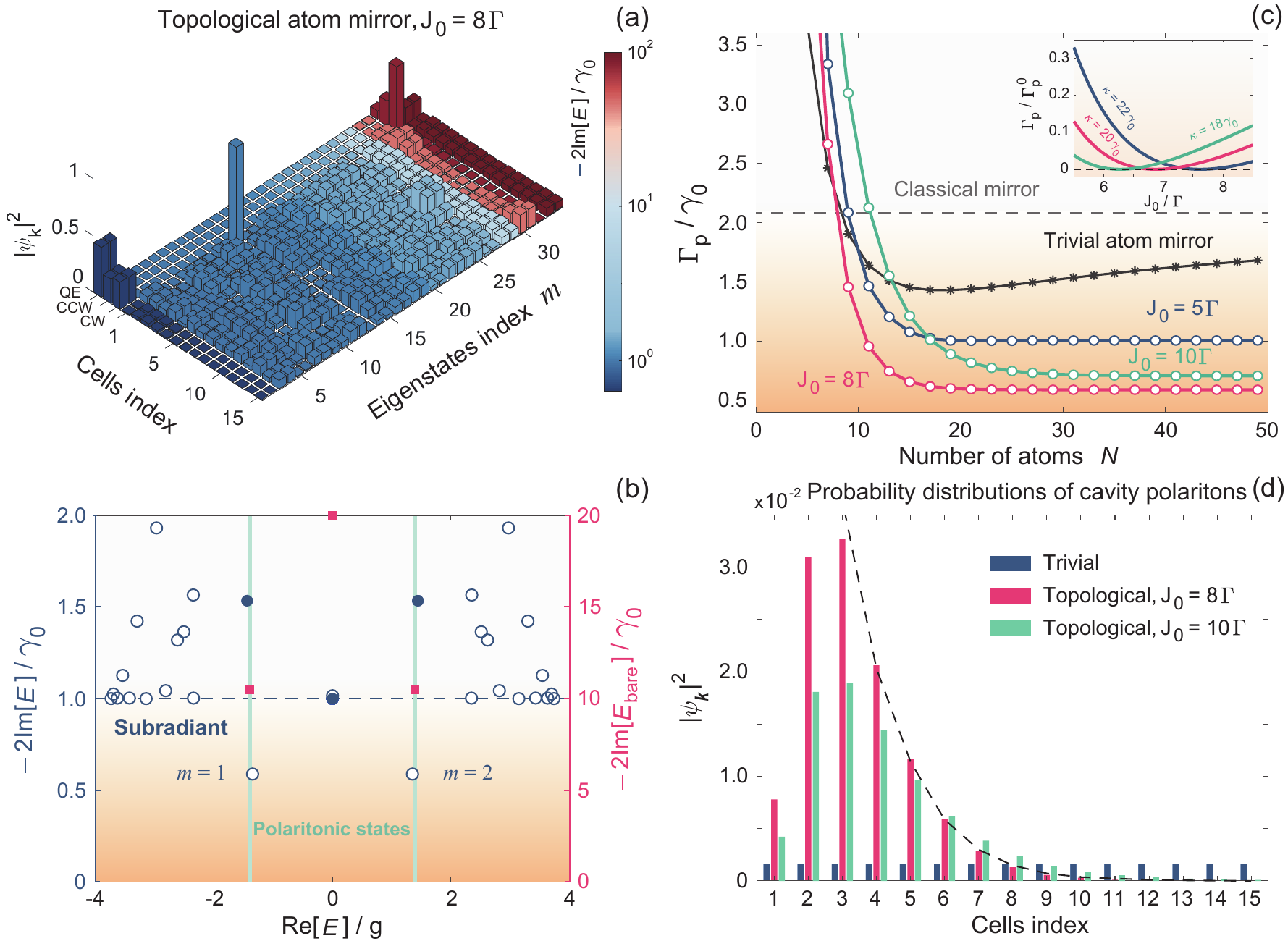}
\caption{(a) Probability distributions of the composed system. Color of each eigenstate gives the decay rate (imaginary eigenenergy). (b) Complex eigenenergies for cavity QED system with topological (blue circles) and trivial atom mirrors (blue dots). The horizontal dashed line indicates the subradiant region and the vertical green thick lines label the polaritonic states. Complex eigenenergies of bare cavity QED system is also shown for comparison (pink squares), which corresponds to the right axis. (c) Decay rate of topological cavity polaritons $\Gamma_p$ versus odd $N$ for various $J_0$. The results of cavity polaritons with trivial atom mirror ($\Gamma^0_p$) and perfect classical mirror (i.e., with unity reflectivity, see Ref. \cite{RN24} for the model in detail) are shown for comparison. The inset shows $\Gamma_p / \Gamma^0_p$ for various $\kappa$ with $J_0=8\Gamma$. (d) Probability distributions of topological edge state-engineered cavity polaritons [$m=1, 2$ in (a)] and trivial cavity polaritons. Dashed black curve is exponential fits to the probability of cavity polaritons versus cells index. Parameters not mentioned are $g=20\gamma_0$, $\kappa=20\gamma_0$, $J_0=8\Gamma$, $\Gamma=5\gamma_0$, $N=31$, $\phi_1=0$ and $\phi=0.3\pi$. }
\label{fig2}
\end{figure*}

\noindent where $\omega_c$ is the frequency of cavity modes, which resonantly couples to QE with strength $g$. $\omega_j$ is the transition frequency of the $j$th atoms and we assume $\omega_j=\omega_c$ unless specially noted. The staggered hoppings $J_j = J_{-}$ ($J_{+}$) for an odd (even) $j$ result in the dimerized interactions between atoms (see schematic presented in Fig. \ref{fig1}). Explicitly, the staggered hoppings can be written as $J_\pm=J_0 [1\pm\mathrm{cos}(\phi)]$, with $J_0$ and $\phi$ being the interaction strength and the tunable parameter of dimerization strength that control the bandgap and localization of edge states, respectively. In absence of dimerized interactions ($J_0 = 0$), the band structure of atom mirror is topologically trivial, which is centrosymmetric with respect to $d=\lambda_0/2$ and plotted in the inset of Fig. \ref{fig1}. The band structure is modified by the dimerized interactions and gives rise to localized edge states in the strong topological regime with $J_0 \gg \gamma_0$, which exhibits the periodicity of $\lambda_0=2\pi/k_0$ in $d$. The inset of Fig. \ref{fig1} also plots the band structure of topological atom mirror with an odd number of sites and $J_0=8\Gamma$, where it shows that a single edge state survives and is isolating from the bulk states due to the presence of energy gap. It also shows that the edge state is exactly protected from the waveguide-mediated interaction for two atom separations, $d=\lambda_0/4$ and $3\lambda_0/4$ (indicated by the vertical dashed line in the inset of Fig. \ref{fig1}), where the coupling between topological atoms is fully dispersive \cite{PRA2017,PRA2022} but no energy shift is observed. This protection stems from the chiral symmetry of SSH chain; however, the topological phases of $d=\lambda_0/4$ and $3\lambda_0/4$ are distinct \cite{RN53}: the former is dissipative while the latter is dissipationless. A brief discussion on topological phase transition can be found in Appendix \ref{ab}. Hereafter, atom mirrors with and without the dimerized interactions are called the topological and trivial atom mirrors, respectively, their interaction with the polaritonic states of strong-coupling cavity QED system yields the topological and non-topological cavity polaritons. In the following study, we focus on the case of $d=3\lambda_0/4$, where the dissipationless edge state can produce prominent anisotropic scattering of photon \cite{RN54}. The coupling of cavity QED system to the dissipationless edge state can suppress the cavity dissipation and results in significant linewidth narrowing of cavity polaritons in both weak- and strong-coupling regimes. 

We consider a single excitation in the composed system, where the subradiant single-photon states hold a great promise for applications related to quantum memory and quantum information storage \cite{RN33,RN35}. To better understand how the topological edge state affects the quantum dynamics and photon transport, we derive the effective Hamiltonian from Eqs. (\ref{eq1})-(\ref{eq6}) under the open boundary condition for atom mirror, which is given by
\begin{equation}
H_{\mathrm{eff}}=H_0^n+H_I+H_{\text {topo }}+H_{\mathrm{vp}}
\end{equation}
with the non-Hermitian free Hamiltonian
\begin{equation}
\begin{aligned}
H_0^n & = \left(\omega_c-i \frac{\kappa}{2}\right) c_{c c w}^{\dagger} c_{c c w}+\left(\omega_c-i \frac{\kappa}{2}\right) c_{c w}^{\dagger} c_{c w} \\ + 
& \left(\omega_c-i \frac{\gamma_0}{2}\right) \sigma_{+}^{(0)} \sigma_{-}^{(0)}+\sum_{j=1}^N\left[\omega_c-i\left(\Gamma+\frac{\gamma_0}{2}\right)\right] \sigma_{+}^{(j)} \sigma_{-}^{(j)}
\end{aligned}
\end{equation}
and the virtual-photon interaction Hamiltonian accounting for the waveguide-mediated long-range hoppings
\begin{equation}
\begin{aligned}
H_{\mathrm{vp}} & =-i \sum_{\substack{j, l=1 \\ j \neq l}}^N \Gamma e^{i\left|\phi_j-\phi_l\right|}\left(\sigma_{+}^{(j)} \sigma_{-}^{(l)}+\sigma_{+}^{(l)} \sigma_{-}^{(j)}\right) \\
& -i \sum_{j=1}^N \sqrt{\kappa \Gamma} e^{i \phi_j}\left(\sigma_{+}^{(j)} c_{c c w}+c_{c w} \sigma_{-}^{(j)}\right)
\end{aligned}
\end{equation}
where the first and second lines characterize the nonlocal interactions between atoms and between the cavity modes and the atoms, respectively. $\phi_j=k_0 x+(j-1)\varphi$ is the effective phase of light propagating from the waveguide-cavity junction to the $j$th atom, where $\varphi=k_0 d$. Due to the open boundaries of the system, we directly diagonalize $H_{\mathrm{eff}}$ to obtain the single-photon band structure and the corresponding eigenstates. For the case of $N=31$ atoms, Fig. \ref{fig2}(a) displays the probability distributions of all eigenstates, which is indexed as $m=1,2,…,34$ by increasingly sorting the decay rates (i.e., $-\mathrm{Im}[E]$, the imaginary parts of eigenenergies $E$) versus system components, including the QE and two cavity modes of cavity QED system and the dimer cells of topological atom mirror. A remarkable feature is that the probability presents a substantial atom content for most of the eigenstates, while concentrates in the cavity QED system for eigenstates labelled $m=1,2$ and $32-34$, i.e., the first two eigenstates with smallest decay rate and the last three with fastest decay. The eigenenergies shown in Fig. \ref{fig2}(b) reveal that the first two eigenstates are essentially the same as the cavity polaritons of bare cavity QED system (i.e., without the atom mirror), but their decay rates are significantly reduced by over an order of magnitude and even smaller than the free-space decay rate $\gamma_0$ of QE, referred to {\it{subradiant cavity polaritons}}. On the contrary, the subradiance cannot be generated for a non-topological cavity polaritons and its decay rate is nearly triple to the topological one. 

Fig. \ref{fig2}(c) compares the decay rates of topological and non-topological cavity polaritons versus the number of atom $N$, where it shows that the non-topological cavity polaritons has the advantage of slow decay with a few atoms; however, the decay rate of topological cavity polaritons rapidly drops with increased $N$ and becomes smaller than that of non-topological cavity polaritons with around 10 atoms. For $J_0=8\Gamma$, 21 atoms are sufficient to reduce the decay rate of topological cavity polaritons to the minimum, and this minimum value is stable as the atom number increases. While the decay rate of non-topological cavity polaritons gradually increases after an optimal $N$ due to the accumulated loss in the system. Fig. \ref{fig2}(c) also indicates that there exists an optimal interaction strength $J_0\approx8\Gamma$, denoted as $J_0^\mathrm{opt}$, corresponding to the smallest decay rate of topological cavity polaritons, which is about $\gamma_0/2$, a half of the decay rate of a bare atom. It implies that the reduction of decay is achieved by suppressing the dissipation of cavity modes, since on resonance the decay rate of cavity polaritons is the average of atom and photon components. Therefore, cavity polaritons can acquire permanent coherence with a QE whose free-space decay vanishes (i.e., $\gamma_0=0$). This claim is confirmed by the results presented in the inset of Fig. \ref{fig2}(c), where it shows that in such a case, the decay of topological cavity polaritons can be completely suppressed for different $\kappa$. It reveals the formation of {\it{bound}} cavity polaritons in a fully open architecture, which is not found with a trivial atom mirror. 

Besides the decay rate, the probability distributions of non-topological and topological cavity polaritons are also distinct, as Fig. \ref{fig2}(d) shows. For non-topological cavity polaritons, the probability is uniformly distributed at each cell of atom mirror due to the translational symmetry of homogeneous chain, while localizes at the left boundary for topological cavity polaritons with $J_0=8\Gamma$ and converges to zero after 10 cells. Furthermore, the probability distributions of topological cavity polaritons manifest the behavior of exponential decay from the left boundary, except for the first two cells since the chiral symmetry is broken in our model. These features indicate the formation of edge state in topological atom mirror and its efficient coupling to the cavity QED system, which is the foundation to realize the topology-engineered cavity polaritons. Similar phenomena are also observed for $J_0=10\Gamma$, but the probability distribution is less concentrated and extended closer to the right boundary with a slow decay. The strong delocalization of probability distributions for a large $J_0$ is a consequence of waveguide-mediated long-range hoppings between topological atoms \cite{QBS,RN64,RN54}. The delocalization tends to populate all topological atoms and leads to the significant decline of reflection when $J_0$ exceeds a critical value where the probability is extended to the last cell of topological atom mirror, as illustrated in Fig. \ref{figs2}(b) of Appendix \ref{ab}. As a result, the delocalization weakens the ability of topological atom mirror in suppressing the leakage of cavity modes. In this situation, more atoms are required to hinder the extension of probability to the right boundary and reduce the decay of topological cavity polaritons. On the other hand, Fig. \ref{fig2}(c) also shows the increased decay of topological cavity polaritons for a small $J_0$. It is attributed to the coupling of topological cavity polaritons to bulk states and yields $J_0^\mathrm{opt}$ for the minimum decay, which we will discuss in the later part of the work.

\subsection{Linewidth narrowing and the enhanced lifetime of subradiant cavity polaritons}
\begin{figure*}[t]
\centering\includegraphics[width=0.96\linewidth]{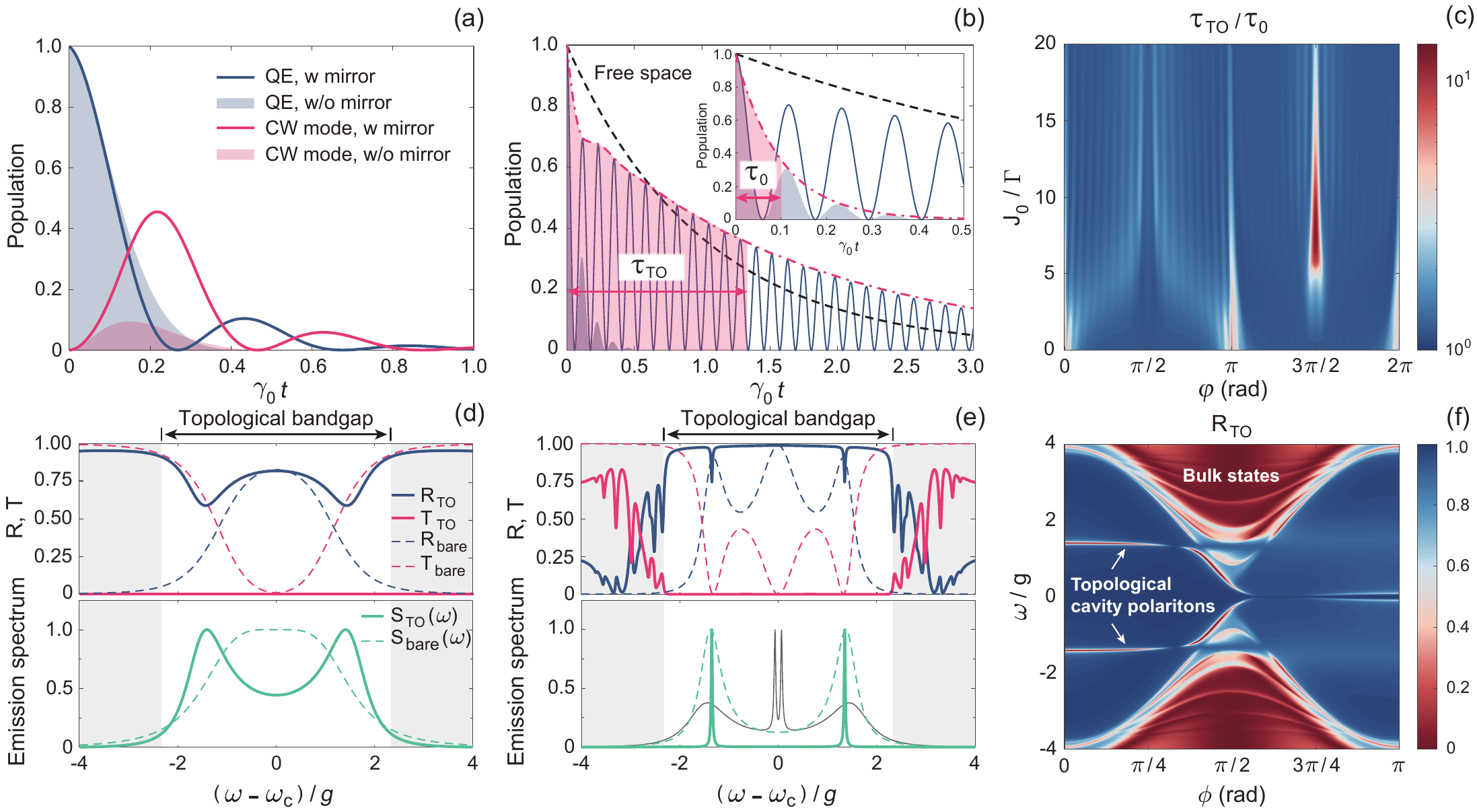}
\caption{(a) and (b) Population dynamics of QE and cavity mode with and without the topological atom mirror in the weak- and strong-coupling regimes, respectively. The inset in (b) shows the short-time dynamics. The lifetime of cavity polaritons is defined as the time that the population decays from $1$ to $e^{-1}$. The black dashed line shows the dynamics of initially excited bare QE in the free space. (c) Topology-enhanced lifetime $\tau_\mathrm{TO}/\tau_0$ versus $\varphi = k_0 d$ in the strong coupling. (d) and (e) Reflection and transmission for left-incident planewave (upper panel) and normalized emission spectrum (lower panel) corresponding to the parameters of (a) and (b), respectively. The non-shaded region indicates the topological bandgap. The light gray line in (e) shows the emission spectrum of QE with $\phi=0.85\pi$. (f) Reflection with topological atom mirror versus $\phi$. Parameters for weak coupling are $g=5\gamma_0$, $\kappa=20\gamma_0$, $J_0=5\Gamma$, $\Gamma=5\gamma_0$, $N=31$, $\phi_1=0$ and $\phi=0.3\pi$. While the strong coupling is $g=20\gamma_0$, $J_0=8\Gamma$ and other parameters remain unchanged. The critical coupling strength for strong coupling is $g_c=\left(\kappa+\gamma_0\right) / 2 \sqrt{2}$ \cite{RN70}. The subscripts 'TO' and 'bare' indicate the spectra with and without the topological atom mirror, respectively.}
\label{fig3}
\end{figure*}

The subradiance of topological cavity polaritons enables to enhance the quantum coherence, demonstrating the features of slow population decay and linewidth narrowing in spectrum. To investigate the quantum dynamics, we derive the equations of motion from the extended cascaded quantum master equation [Eqs. (\ref{eq1})-(\ref{eq6})]
\begin{equation}
\frac{d}{d t} \tilde{\sigma}_{-}^{(0)}=-\frac{\gamma_0}{2} \tilde{\sigma}_{-}^{(0)}-i g\left(\tilde{c}_{c w}+\tilde{c}_{c c w}\right)
\end{equation}
\begin{equation}
\frac{d}{d t} \tilde{c}_{c c w}=-\frac{\kappa}{2} \tilde{c}_{c c w}-i g \tilde{\sigma}_{-}^{(0)}
\end{equation}
\begin{equation}
\frac{d}{d t} \tilde{c}_{c w}=-\frac{\kappa}{2} \tilde{c}_{c w}-i g \tilde{\sigma}_{-}^{(0)}-\sqrt{\kappa \Gamma} \sum_{j=1}^N \tilde{\sigma}_{-}^{(j)} e^{i \phi_j}
\end{equation}
\begin{equation}
\begin{aligned}
\frac{d}{d t} \tilde{\sigma}_{-}^{(j)}= & -\frac{\gamma_0}{2} \tilde{\sigma}_{-}^{(j)}-\sqrt{\kappa \Gamma} e^{i \phi_j} \tilde{c}_{c c w}-\Gamma \sum_{l=1}^N \tilde{\sigma}_{-}^{(l)} e^{i\left|\phi_j -\phi_l\right|} \\
& -J_{j-1} \tilde{\sigma}_{-}^{(j-1)}-J_{j+1} \tilde{\sigma}_{-}^{(j+1)}
\end{aligned}
\end{equation}
where the substitution $\langle O\rangle=\operatorname{Tr}[O \rho]=\tilde{O} e^{-i \omega_c t}$ is applied and the single-photon approximation $\left[\sigma_{+}, \sigma_{-}\right]=\sigma_z=-1$ is imposed. Figs. \ref{fig3}(a) and (b) plot the population dynamics $\left|\tilde{\sigma}_{-}^{(0)}\right|^2$ of an initially excited QE with parameters of the weak- and strong-coupling regimes for bare cavity QED system, respectively. By reducing the losses associated with the leakage of cavity modes through the topological atom mirror, a weak-coupling cavity QED system can enter into the strong-coupling regime, which is evidenced by Rabi oscillation in the population dynamics of both QE and cavity modes and shown in Fig. \ref{fig3}(a). While for a bare cavity QED system already in the strong-coupling regime, Fig. \ref{fig3}(b) shows that the period of Rabi oscillation is almost unchanged after introducing the topological atom mirror, while its decay is strongly suppressed and even slower than a bare QE in the free space. It implies that the coupling strengths of QE-cavity interaction are comparable in two configurations but the linewidth of cavity polaritons is significantly narrowed. As a consequence, the lifetime of cavity polaritons can be prolonged by over an order of magnitude, see the lifetime enhancement $\tau_\mathrm{TO}/\tau_0$ shown in Figs. \ref{fig3}(b) and (c), where $\tau_\mathrm{TO}$ and $\tau_0$ are the lifetimes of topological and non-topological cavity polaritons, respectively. We find that for topological cavity polaritons, $\tau_\mathrm{TO}$ allows more than 11 cycles of energy exchange between the QE and the cavity, while the non-topological cavity polaritons cannot accomplish a complete period of Rabi oscillation within $\tau_0$. We also find that $\tau_\mathrm{TO}/\tau_0$ depends on the choice of $\varphi$ and there is a narrow window of $\varphi$ for significant enhancement of lifetime, offering a degree of tuning tolerance to fabrication errors and experimental uncertainties. The maximum enhancement of lifetime, or equivalently, the greatest linewidth narrowing, is found at $\varphi=3\pi/2$.

To observe the phenomenon of linewidth narrowing associated with the subradiance of topological cavity polaritons, we calculate the spectra of the composed system for two excitation configurations, the reflection and transmission for left-incident planewave and the fluorescence of QE addressed through the free space, which are both experimentally relevant. For the first configuration, the dynamics of system can be written in a compacted form (see Appendix \ref{aaa} and also Ref. \cite{LZY} for detailed derivation)
\begin{equation}\label{eq14}
\frac{d }{d t}\langle\mathbf{s}(t)\rangle=-i V^{-1} E V \langle\mathbf{s}(t)\rangle-P_{i n}
\end{equation}
with
\begin{equation}\label{eq15}
\mathbf{s}(t)=\left[\sigma_{-}^{(0)}(t), c_{c c w}(t), c_{c w}(t), \sigma_{-}^{(1)}, \cdots, \sigma_{-}^{(N)}(t)\right]^T
\end{equation}
and
\begin{equation}\label{eq16}
P_{i n}=\left[0, \sqrt{\kappa}, 0, \sqrt{\Gamma} e^{i \phi_1}, \cdots, \sqrt{\Gamma} e^{i \phi_N}\right]^T a_{i n}
\end{equation}
where $E$ and $V$ are the eigenvalues and the corresponding left eigenvectors of $H_\mathrm{eff}$. $a_\mathrm{in}$ is the amplitude of input field. The solution in the frequency domain is given by
\begin{equation}\label{eq17}
\mathbf{s}(\Delta)=i V^{-1}(\Delta I-E)^{-1} V P_{i n}
\end{equation}
where $\Delta=\omega-\omega_c$ is the frequency detuning. The output fields of waveguide are given by
\begin{equation}\label{eq18}
a_{\text {out }}=R_{\text {out }} \mathbf{s}(\Delta)
\end{equation}
\begin{equation}\label{eq19}
b_{\text {out }}=a_{in}e^{i \phi_N}+T_{\text {out }} \mathbf{s}(\Delta)
\end{equation}
with 
\begin{equation}\label{eq20}
R_{\text {out }}=\left[0,0, \sqrt{\kappa} e^{i \phi_N}, \sqrt{\Gamma} e^{i(N-1) \varphi}, \cdots, \sqrt{\Gamma} e^{i \varphi}, \sqrt{\Gamma}\right]^T
\end{equation}
\begin{equation}\label{eq21}
T_{\text {out }}=\left[0, \sqrt{\kappa} e^{i \phi_N}, 0, \sqrt{\Gamma} e^{i(N-1) \varphi}, \cdots, \sqrt{\Gamma} e^{i \varphi}, \sqrt{\Gamma}\right]^T
\end{equation}
Subsequently, we can obtain the reflection and transmission spectra as $R(\Delta)=a_{\text {out}}^{\dagger}(\Delta) a_{\text {out}}(\Delta) /\left|a_{\text {in}}\right|^2$ and $T(\Delta)=b_{\text {out}}^{\dagger}(\Delta) b_{\text {out}}(\Delta) /\left|a_{\text {in}}\right|^2$, respectively. Eqs.(\ref{eq18})-(\ref{eq21}) indicate that in this configuration, the pump photons can interfere with the scattering photons. While for the configuration of fluorescence, only the photons emitted by QE are detected. The emission spectrum of QE is defined as $S(\omega)=\lim _{t \rightarrow \infty} \operatorname{Re}\left[\int_0^{\infty} d \tau\left\langle\sigma_{+}^{(0)}(t) \sigma_{-}^{(0)}(t+\tau)\right\rangle e^{i \omega \tau}\right]$, where the two-time correlation function $\left\langle\sigma_{+}^{(0)}(t) \sigma_{-}^{(0)}(t+\tau)\right\rangle$ can be obtained by using the quantum regression theorem \cite{RN2,RN24}. The equation for two-time correlation functions is as follows: 
\begin{equation}
\frac{d }{d \tau}\mathbf{c}(\tau)=-i V^{-1} E V \mathbf{c}(\tau)
\end{equation}
where $\mathbf{c}(\tau)=\left[\left\langle\sigma_{+}^{(0)}(0) \mathbf{s}(\tau)\right\rangle\right]^T$, with $\mathbf{s}(\tau)$ given in Eq. (\ref{eq15}). With initial condition $c_0=[1,0, \cdots, 0]^T$, we have $\mathbf{c}(\omega)=i V^{-1}(\omega I-E)^{-1} V c_0$, which yields the solution of $\left\langle\sigma_{+}^{(0)} \sigma_{-}^{(0)}(\omega)\right\rangle$ in the frequency domain. On the other hand, the emission spectrum of QE can be expressed as $S(\omega)=\operatorname{Re}\left[\left\langle\sigma_{+}^{(0)} \sigma_{-}^{(0)}(\omega)\right\rangle\right]$ by using the Fourier transform relations. We thus can obtain the emission spectrum of QE by substituting $\left\langle\sigma_{+}^{(0)} \sigma_{-}^{(0)}(\omega)\right\rangle$ into $S(\omega)$. 

For the weak-coupling cavity QED system corresponding to Fig. \ref{fig3}(a), no spectral splitting is observed in both the transmission/reflection spectrum and the emission spectrum of QE for the bare cavity-QED without topological atom mirror, which are shown in the dashed curves in Fig. 3(d). However, we note that the emission spectrum of QE is obviously broaden, which is the superposition of two eigenmodes and it can be called as ‘dark’ strong coupling \cite{PRBLLW}. In contrast, the topological atom mirror brings the system into the strong-coupling regime, characterized by resolvable Rabi splitting with a width of $\sim 2 \sqrt{2} g$ seen in both the reflection spectrum and the emission spectrum of QE. In this case, the transmission is approximately zero since the edge state is localized at the left boundary of atom mirror [see Fig. \ref{fig2}(c) and Fig. \ref{figs2}(c) in Appendix \ref{ab}]. While for a strong-coupling cavity QED system, Fig. \ref{fig3}(e) shows that the Rabi peaks corresponding to topological cavity polaritons exhibit extremely sharp linewidth, which can be apparently observed in both the reflection spectrum and the emission spectrum of QE. Besides the Rabi splitting, multiple resonances located at the left and right sides of the reflection and transmission spectra are reminiscent of bulk states. 

The topological bandgap separates the edge states and bulk states and plays a central role in determining the optical response of atom mirror. The gap of topological band can be controlled by the parameter $\phi$. It is interesting to see the variation of reflection spectrum by tuning $\phi$, which can reveal how the change of underlying band structure alters the decay of topological cavity polaritons. As the reflection spectrum of Fig. \ref{fig3}(f) shows, the locations of upper and lower bands can be clearly identified through the boundary where the reflection suddenly drops. Two bands are symmetrical with respect to $\phi=\pi/2$, but the linewidth of topological cavity polaritons is obviously increased as $\phi$ goes though $\pi/2$. In the parameter range of $\pi / 2<\phi \leq \pi$, the coupling of strong-coupling cavity QED system to topological edge state slightly broadens instead of narrowing the linewidth of polaritonic states, as the light gray line in the lower panel of Fig. \ref{fig3}(e) shows; meanwhile, there are two hybrid edge modes with a finite gap around the zero energy, similar to a SSH chain with even sites \cite{RN47,RN64}. The dramatic change of linewidth results from the opposite localization of edge states, which localize at the left (right) boundary of atom mirror for $0 \leq \phi \leq \pi / 2$ ($\pi / 2<\phi \leq \pi$), see the examples shown in Figs. \ref{figs2}(c) and (d) in Appendix \ref{ab}. The results presented in Fig. \ref{fig3}(f) demonstrate the capacity of topological edge states in efficiently tuning the lifetime and linewidth of cavity polaritons at the single-quantum level. The emission spectrum of QE shown in Fig. \ref{figs3}(a) of Appendix \ref{ad} demonstrates the similar features observed in the reflection spectrum, but the signal of multiple resonances corresponding to bulk states are weak. Therefore, it is preferable to investigate the properties of topological cavity polaritons through the fluorescence of QE.

\begin{figure*}[t]
\centering\includegraphics[width=\linewidth]{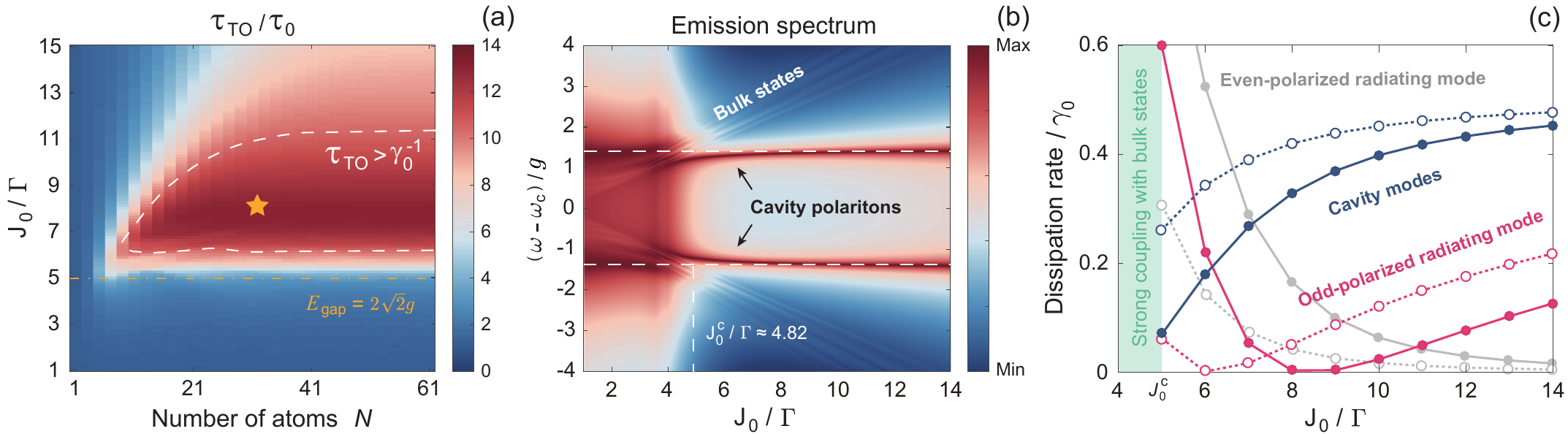}
\caption{(a) Topology-enhanced lifetime of cavity polaritons $\tau_\mathrm{TO}/\tau_0$ versus the number of atoms $N$ and the interaction strength $J_0$. Orange dashed dotted line indicates the critical $J_0$ (i.e., $J_0^c$) for topological bandgap with a width of $E_{\text {gap }}=2 \sqrt{2} g$. Dashed white line surrounds the parameters range of $\tau_\mathrm{TO}>\gamma_0^{-1}$. Orange star denotes $\tau_\mathrm{TO}/\tau_0$ for dynamics shown in Fig. \ref{fig3}(b). (b) Emission spectrum of QE versus $J_0$. The horizontal and vertical white dashed lines indicate the locations of cavity polaritons and $J_0^c$, respectively. (c) Dissipation from topological cavity polaritons to two cavity modes (blue lines) and the even- and odd-polarized radiating modes (light gray and red lines) for $\phi=0.3\pi$ (solid lines with dots) and $0.2\pi$ (dashed lines with circles). Parameters not mentioned are the same as Fig. \ref{fig3}(b). }
\label{fig4}
\end{figure*}

$J_0$ is another important parameter that determines the topological bandgap, and hence the lifetime $\tau_\mathrm{TO}$ exhibits strong dependence on $J_0$. Fig. \ref{fig4}(a) plots the lifetime enhancement $\tau_\mathrm{TO}/\tau_0$ as the function of interaction strength $J_0$ and the number of atoms $N$, where it shows that $\tau_\mathrm{TO}>\gamma_0^{-1}$ can be achieved in a wide range of $J_0$ with a few tens of atoms in mirror. Remarkably, $\tau_\mathrm{TO}/\tau_0$ demonstrates an abrupt increase at a critical interaction strength $J_0^c$ regardless of $N$. This phenomenon can be understood by inspecting the emission spectrum of QE versus $J_0$, as shown in Fig. \ref{fig4}(b). We can see that the significant linewidth narrowing of Rabi peaks occurs at $J_0^c$ corresponding to the topological bandgap slightly larger than the width of Rabi splitting, i.e., $J_0^{\mathrm{c}} \gtrsim g / \sqrt{2} \cos (\varphi)$. The reason is that for $J_0>J_0^c$, the cavity polaritons are detuned from the superradiant bulk states and the corresponding dissipation is strongly suppressed, resulting in the significant enhancement of lifetime. However, a large $J_0$ is not always beneficial for enhancing the lifetime. As is seen in Fig. \ref{fig4}(a) and discussed earlier, there exists an optimal interaction strength $J_0^\mathrm{opt}$ for maximal $\tau_\mathrm{TO}/\tau_0$ as a consequence of the tradeoff between the delocalization of edge states and the dissipation induced by bulk states. We also notice that the anticrossing behavior can be observed at $J_0\sim4\Gamma$ [see Fig. \ref{figs3}(c) in Appendix \ref{ad} for a closeup and further discussion], implying the strong coupling between the topological cavity polaritons and the bulk states.


\begin{figure*}[t]
\centering\includegraphics[width=\linewidth]{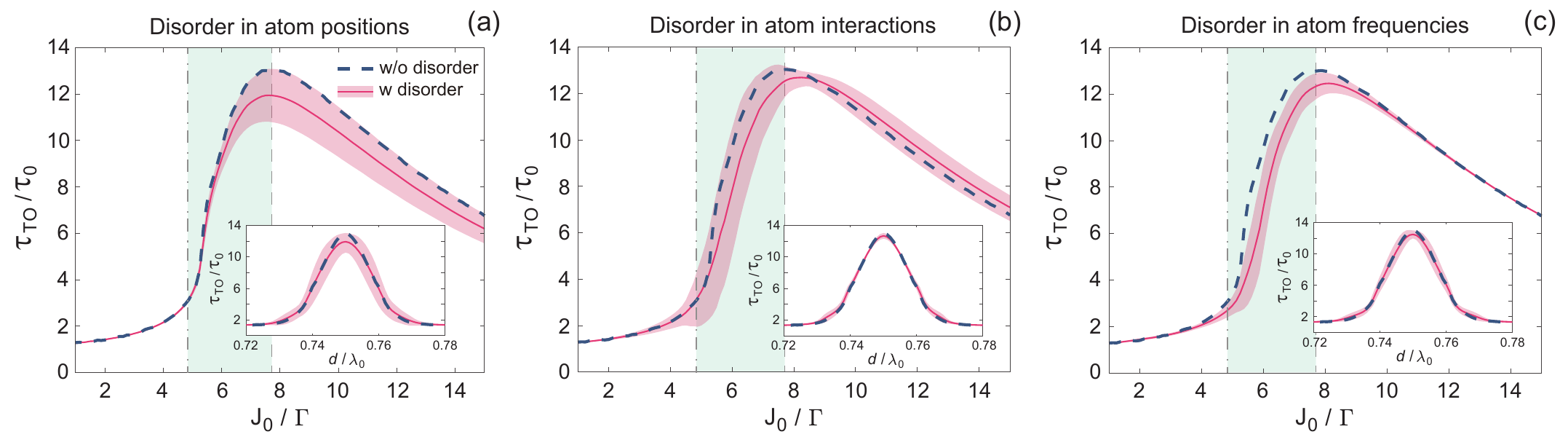}
\caption{Topologically robust enhancement of lifetime $\tau_\mathrm{TO}/\tau_0$ versus interaction strength $J_0$ with disorder in atom positions (a), atom interactions (b), and atom frequencies (c). The disorder ranges are $\Delta d_j \in [-0.02 d, 0.02 d]$ for atom positions, $\Delta J_j \in [-0.2 J_0, 0.2 J_0]$ for atom interactions, and $\Delta\omega_j \in [-g / \sqrt{2}, g / \sqrt{2}]$ for atom frequencies. The pink solid line represents the mean value averaged over 100 random realizations of disordered topological atom mirror, with pink shaded area indicating the standard deviation. The green shaded region indicates the parameter range of $J_0^c \leq J_0 \leq J_0^\mathrm{opt}$, where the vertical dashed dotted line and dashed line label $J_0^c$ and $J_0^\mathrm{opt}$, respectively. The insets show $\tau_\mathrm{TO}/\tau_0$ at $J_0=J_0^\mathrm{opt}$ versus atom spacing $d$. Other parameters are the same as Fig. \ref{fig3}(b). }
\label{fig5}
\end{figure*}

To shed insights into the dissipative properties of topological cavity polaritons, we diagonalize the Lindblad operator [Eq. (\ref{eq2})] to obtain the dissipative matrix $\gamma=\sum_m \chi_m\left|v_m\right\rangle\left\langle v_m\right|$, with $\chi_m$ being the dissipation spectrum, which is shown in Fig. \ref{figs4}(a) (see Appendix \ref{ae} for more details). Four dissipative modes are found to have large dissipation rate, which are two polarized radiating modes corresponding to the even and odd sites in topological atom mirror and other two are related to cavity modes, as we see from the wave function shown in Fig. \ref{figs4}(b) of Appendix \ref{ae}. In our model, the dissipation of odd-polarized radiating mode is greater than that of even-polarized radiating mode due to the odd number of atoms. The dissipation rate $\Gamma_{ \pm}$ from topological cavity polaritons to environment is given by the overlap between the polaritonic states and the radiating modes in the Lindblad operator, which is evaluated as \cite{RN53}
\begin{equation}
\Gamma_{ \pm}=\left\langle\psi_{ \pm}|\gamma| \psi_{ \pm}\right\rangle=\sum_m \chi_m\left\langle\psi_{ \pm}| \chi_m\right\rangle^2
\end{equation}
where $\left|\psi_{ \pm}\right\rangle$ is the eigenstate of $\operatorname{Re}\left[H_{\text {eff }}\right]$ corresponding to the cavity polaritons. The results are plotted in Fig. \ref{fig4}(c), where it shows that the dissipation rate of odd-polarized radiating mode approaches to zero around $J_0^\mathrm{opt}$, while the dissipation rate contributed by the even-polarized radiating mode (cavity modes) is monotonically decreasing (increasing) with increased $J_0$. We also find that the dissipation of radiating modes dramatically increases as $J_0$ approaches to $J_0^c$. In addition, a small $\phi$ that produces a large topological bandgap [see Fig. \ref{fig3}(f)] can reduce the dissipation rate of radiating modes. These results suggest that the system energy mainly dissipates from bulk states to environment in parameter range of $J_0<J_0^\mathrm{opt}$. While for $J_0>J_0^\mathrm{opt}$, the dissipation of cavity modes is dominated over the radiating modes and as a result, the minimum $\Gamma_{ \pm}$ achieves around $J_0$ corresponding to zero dissipation rate ($\sim10^{-3}\gamma_0$) of odd-polarized radiating mode. 

\subsection{Robustness against the disorder in topological atom mirror}

In practice, the perturbations on system and imperfections of structure are inevitable. In this subsection, we investigate the impact of local disorder on the enhancement of lifetime for topological cavity polaritons. In Fig. \ref{fig5}(a), we plot $\tau_\mathrm{TO}/\tau_0$ with disordered positions for topological atoms, where the position of the $j$th atom is $d_j+\Delta d_j$, with $-0.02 d \leq \Delta d_j \leq 0.02 d$. It shows that the disorder in atom positions has small impact on the lifetime for $J_0<J_0^\mathrm{opt}$, especially when $\tau_\mathrm{TO}/\tau_0<10$. Though it affects the lifetime in a negative manner, we find that $\tau_\mathrm{TO}/\tau_0>10$ can still be obtained around $J_0^\mathrm{opt}$. Different from the disorder in atom positions, the positive impact on lifetime is observed for disorder in interaction strengths with moderate disorder $-0.2 J_0 \leq \Delta J_j \leq 0.2 J_0$ and $J_0>J_0^\mathrm{opt}$, as Fig. \ref{fig5}(b) shows. In this case, the interaction strength between the $j$th and ($j+1$)th atoms is given by $J_j+\Delta J_j$. The inset of Fig. \ref{fig5}(b) displays that $\tau_\mathrm{TO}/\tau_0$ at $J_0=J_0^\mathrm{opt}$ manifests high robustness against the disorder in interaction strengths for various atom spacing. The results presented in Figs. \ref{fig5}(a) and (b) indicate that the disorder in atom positions has greater impact on the lifetime of topological cavity polaritons, as we can see that the lifetime variation of $2\%$ disorder in atom positions is comparable and even slightly larger than that of $20\%$ disorder in atom interactions around $J_0^\mathrm{opt}$. It is because the impact of disorder in atom positions is nonlocal, which affects the coupling between the disordered atom and all other atoms through the waveguide-mediated long-range hoppings. On the contrary, the disorder in atom interactions is local perturbation, which only alters the coupling between neighboring sites. Therefore, the lifetime enhancement manifests higher robustness against disorder in atom interactions. As for disorder in atom frequencies, its main effect is on the energies of edge and bulk states, thus the impact on the lifetime of topological cavity polaritons is not obvious if the topological bandgap is sufficiently large. Fig. \ref{fig5}(c) shows the lifetime enhancement in presence of disorder in atom frequencies, where the strong disorder strength is considered. The frequency of the $j$th atom is randomly distributed in range of $\omega_j \in [\omega_c-g / \sqrt{2}, \omega_c+g / \sqrt{2}]$, where the maximal disorder strength is a half of the width of Rabi splitting. We can see that similar to Fig. \ref{fig5}(b), disorder in atom frequencies also begins to have noticeable effect around $J_0^c$ (vertical dashed dotted line), but in this case the lifetime of topological cavity polaritons is less sensitive to disorder when $J_0>J_0^\mathrm{opt}$ (vertical dashed line) as expected. Therefore, it implies that disorder in atom interactions and frequencies mainly affect the edge states and bulk states of topological atom mirror, respectively. We conclude from Fig. \ref{fig5} that the presence of moderate disorder in atom mirror will not severely spoil the enhanced lifetime of topological cavity polaritons. 


It should be emphasized that the parameters used in this work are attainable in nanophotonic platform with semiconductor QEs. Taking InGaAs quantum dots for an example, the intrinsic decay is $\gamma_0 \approx 10 \mu \mathrm{eV}$ at cryogenic temperatures \cite{RN71}. For the strong-coupling regime under investigation, $\kappa=20 \gamma_0$ corresponds to a cavity with $Q$ factor of $\sim 6 \times 10^3$. The QE-cavity coupling strength $g=20 \gamma_0=200 \mu \mathrm{eV}$ can be obtained, for instance, in a WGM microdisk with $3\mu$m radius \cite{RN25,RN38}. For quantum dot arrays, the switchable coupling between two neighboring sites is usually provided by the tunnel barrier of electrostatic potential, which can be tuned through control gate \cite{RN50,RNnpjQI,RNgate}. Very recently, the experimental realization of SSH chain based on ten semiconductor QEs with tunable interaction strengths has been reported \cite{RN50}. With state-of-art technology, the precision of positioning a QE can reach $\sim15$nm \cite{RN72}, which is less than $2\%$ compared to the emission wavelength of QE. As the results of Fig. \ref{fig5}(a) indicates, such experimental uncertainties has limited impact on the lifetime of topological cavity polaritons. In addition, for the case of single-photon excitation as we study in this work, the topological atom mirror can be replaced by cavity counterpart \cite{RN59}, which is a more feasible experimental configuration to tune the system parameters. Besides the solid-state QEs, the technology of optical tweezers has already been applied to construct one- and two-dimensional atom arrays with the number of cold atoms upto 200 \cite{RN73,RN74,RN75}. Alternatively, the cavity-magnon systems \cite{RN17} and superconducting circuits \cite{RN35,RN49} are also promising candidates to implement topological atom mirror for the advance in realizing multi-atom interactions over extended distances. Therefore, the considerable enhancement of lifetime by over an order of magnitude predicted here is achievable for cavity polaritons in diverse quantum systems.

\section{Conclusion}
In summary, we propose a scheme for narrowing the linewidth of cavity polaritons combined with robustness by coupling a one-dimensional topological atom mirror to the cavity QED system based on WGM resonator. The cavity polaritons become subradiant, with a linewidth smaller than that of a single QE through the coupling of cavity mode to edge states in dissipationless topological phase. Accordingly, the lifetime can be improved by over an order of magnitude. The subradiance of cavity polaritons are protected by the topological bandgap and hence can survive in the disordered atom mirror.

Our architecture exhibits prominent advantages in at least three aspects. Firstly, the maximal enhancement of lifetime is achieved in a cavity with moderate $Q$ factor of $10^3-10^4$, which gets rid of the drawback of poor excitation and collection efficiencies in conventional approach that reduces the linewidth by the use of a high-$Q$ cavity. This feature combined with the openness of semi-infinite waveguide benefits the practical applications. Importantly, several unit cells, typically $10-20$ atoms, are sufficient to narrow the linewidth of cavity polaritons to a value comparable to a single QE in the free space. Topological atom mirror of this scale has been demonstrated with state-of-art technology of nanofabrication. Last but not least, the property of topological protection empowers the subradiant cavity polaritons to have high tolerance for fabrication imperfections and experimental uncertainties. Moving forward, future endeavors can devote to explore the effects of coherent time-delayed feedback on lifetime enhancement \cite{RNFB}, or conceive the scheme of {\it{in situ}} and dynamical topological manipulation of quantum states \cite{RNnpjQI}. Therefore, our scheme offers a promising platform for exploring topological quantum optics and may be potentially used for long-time storage of quantum states in experiments, which is crucial to push quantum technologies toward practical applications.

\begin{acknowledgments}
Y.W. Lu acknowledges the support of National Natural Science Foundation of China (Grant Nos. 62205061, 12274192). Z. Liao acknowledges the support of National Key R\&D Program of China (Grant No. 2021YFA1400800), the Guangdong Basic and Applied Basic Research Foundation (Grant No. 2023B1515040023) and the Natural Science Foundations of Guangdong (Grant No. 2021A1515010039).
\end{acknowledgments}

\appendix

\begin{widetext}
\section{Derivation of the extended cascaded quantum master equation}\label{aa}

We derive the extended cascaded quantum master equation [Eqs. (\ref{eq1})-(\ref{eq2})] by tracing out the degrees of freedom of the waveguide. The system Hamiltonian including the waveguide modes is given by ($\hbar = 1$)
\begin{equation}
H_S=H+H_w+H_{s w}
\end{equation}
\noindent where $H=H_0+H_I+H_\mathrm{topo}$ is given in Eqs. (\ref{eq3})-(\ref{eq6}). $H_w$ is the free Hamiltonian of waveguide
\begin{equation}
H_w=\sum_{\lambda=R, L} \int d \omega \omega b_\lambda^{\dagger} b_\lambda
\end{equation}
and $H_{sw}$ is the interaction Hamiltonian that describes the cavity-waveguide and atom-waveguide interactions
\begin{equation}
H_{s w}=i \sum_{\lambda=R, L} \int d \omega \sqrt{\frac{\kappa_\lambda}{2 \pi}} b_\lambda^{\dagger} e^{-i k_\lambda x_0} c_\lambda+i \sum_{\lambda=R, L} \sum_{j=1}^N \int d \omega \sqrt{\frac{\gamma_\lambda}{2 \pi}} b_\lambda^{\dagger} e^{-i k_\lambda x_j} \sigma_{-}^{(j)}+H . c .
\end{equation}
where $b_L$ ($b_R$) is the bosonic annihilation operator of the left-propagating (right-propagating) waveguide mode with frequency $\omega$ and wave vector $k_R=-k_L=k_0\equiv\omega_c/v$ with $v$ being the group velocity. Note that for the sake of convenience, we have used the notions $c_R$=$c_{ccw}$ and $c_L=c_{cw}$ since the CCW and CW modes are coupled to the right- and left-propagating guided modes, respectively. $x_0$ is the location of cavity-waveguide junction and $x_j$ indicates the location of $j$th atom that couples to the waveguide. Applying the transformation $\widetilde{H}=U H U^{\dagger}-i d U / d t U^{\dagger}$ with $U=\exp \left[i\left(\omega_c \sum_{\lambda=R, L} c_\lambda^{\dagger} c_\lambda+\sum_{j=1}^N \omega_j \sigma_{+}^{(j)} \sigma_{-}^{(j)}+\sum_{\lambda=R, L} \int d \omega \omega b_\lambda^{\dagger} b_\lambda\right)t\right]$, we have
\begin{equation}
\widetilde{H}_{s w}(t)=i \sum_{\lambda=R, L}\left[\int d \omega \sqrt{\frac{\kappa_\lambda}{2 \pi}} b_\lambda^{\dagger} e^{i\left(\omega-\omega_c\right) t} e^{-i \omega x_0 / v} c_\lambda+\sum_{j=1}^N \int d \omega \sqrt{\frac{\gamma_\lambda}{2 \pi}} b_\lambda^{\dagger} e^{i\left(\omega-\omega_j\right) t} e^{-i \omega x_j / v} \sigma_{-}^{(j)}\right]+H . c .
\end{equation}
The equation of $b_\lambda$ can be obtained from the Heisenberg equation
\begin{equation}\label{eqa5}
\frac{d}{d t} b_\lambda(t)=\sqrt{\frac{\kappa_\lambda}{2 \pi}} c_\lambda(t) e^{i\left(\omega-\omega_c\right) t} e^{-i \omega x_0 / v}+\sum_{j=1}^N \sqrt{\frac{\gamma_\lambda}{2 \pi}} \sigma_{-}^{(j)}(t) e^{i\left(\omega-\omega_j\right) t} e^{-i \omega x_j / v}
\end{equation}
Formally integrating the above equation, we have
\begin{equation}\label{eqa6}
b_\lambda(t)=\int_0^t d \tau \sqrt{\frac{\kappa_\lambda}{2 \pi}} c_\lambda(\tau) e^{i\left(\omega-\omega_c\right) \tau} e^{-i \omega x_0 / v}+\sum_{j=1}^N \int_0^t d \tau \sqrt{\frac{\gamma_\lambda}{2 \pi}} \sigma_{-}^{(j)}(\tau) e^{i\left(\omega-\omega_j\right) \tau} e^{-i \omega x_j / v}
\end{equation}
where the initial condition $b_\lambda (0)=0$ is imposed since the waveguide is in the vacuum state. On the other hand, the equation of motion of arbitrary operator $O$ reads 
\begin{equation}\label{eqa7}
\begin{aligned}
\frac{d}{d t} O(t) & =\sum_{\lambda=R, L} \int d \omega \sqrt{\frac{\kappa_\lambda}{2 \pi}}\left\{b_\lambda^{\dagger}(t) e^{i\left(\omega-\omega_c\right) t} e^{-i \omega x_0 / v}\left[O(t), c_j(t)\right]-\left[O(t), c_j^{\dagger}(t)\right] b_\lambda(t) e^{-i\left(\omega-\omega_c\right) t} e^{i \omega x_0 / v}\right\} \\
& +\sum_{\lambda=R, L} \sum_{j=1}^N \int d \omega \sqrt{\frac{\gamma_\lambda}{2 \pi}}\left\{b_\lambda^{\dagger}(t) e^{i\left(\omega-\omega_j\right) t} e^{-i \omega x_j / v}\left[O(t), \sigma_{-}^{(j)}(t)\right]\right. \left.-\left[O(t), \sigma_{+}^{(j)}(t)\right] b_\lambda(t) e^{-i\left(\omega-\omega_j\right) t} e^{i \omega x_j / v}\right\}
\end{aligned}
\end{equation}
Substituting $b_\lambda (t)$ into the above equation, we obtain 
\begin{equation}\label{eqa8}
\begin{aligned}
\frac{d}{d t} O(t) & =\sum_{\lambda=R, L}\int_0^t d \tau \int d \omega\left\{\left[\frac{\kappa_\lambda}{2 \pi} c_\lambda^{\dagger}(\tau)+\sum_{j=1}^N \frac{\sqrt{\kappa_\lambda \gamma_\lambda}}{2 \pi} \sigma_{+}^{(j)}(\tau) e^{-i \omega x_{0 j} / v}\right]\left[O(t), c_\lambda(t)\right] e^{i\left(\omega-\omega_c\right)(t-\tau)}\right. \\
& \left.-\left[O(t), c_\lambda^{\dagger}(t)\right]\left[\frac{\kappa_\lambda}{2 \pi} c_\lambda(\tau)+\sum_{j=1}^N \frac{\sqrt{\kappa_\lambda \gamma_\lambda}}{2 \pi} \sigma_{-}^{(j)}(\tau) e^{i \omega x_{0 j} / v}\right] e^{-i\left(\omega-\omega_c\right)(t-\tau)}\right\} \\
& +\sum_{\lambda=R, L} \sum_{j=1}^N \int_0^t d \tau \int d \omega\left\{\left[\frac{\sqrt{\kappa_\lambda \gamma_\lambda}}{2 \pi} c_\lambda^{\dagger}(\tau) e^{-i \omega x_{j 0} / v}\right.\right. \left.\left.+\sum_{l=1}^N \frac{\gamma_\lambda}{2 \pi} \sigma_{+}^{(l)}(\tau) e^{-i \omega x_{j l} / v}\right]\left[O(t), \sigma_{-}^{(j)}(t)\right] e^{i\left(\omega-\omega_c\right)(t-\tau)}\right] \\
& \left.-\left[O(t), \sigma_{+}^{(j)}(t)\right]\left[\frac{\sqrt{\kappa_\lambda \gamma_\lambda}}{2 \pi} c_\lambda(\tau) e^{i \omega x_{j 0} / v}+\sum_{l=1}^N \frac{\gamma_\lambda}{2 \pi} \sigma_{-}^{(l)}(\tau) e^{i \omega x_{j l} / v}\right] e^{-i\left(\omega-\omega_c\right)(t-\tau)}\right\}
\end{aligned}
\end{equation}
where $x_{j0}=x_j-x_0$ and $x_{jl}=x_j-x_l$. We perform the Markov approximation by assuming the time delay $x_{jl}/v$ between the atoms and $x_{j0}$ between the cavity modes and the atoms are sufficiently small and can be neglected. Therefore, we have
\begin{equation}\label{eqa9}
\frac{\kappa_\lambda}{2 \pi} \int_0^t d \tau \int d \omega e^{i\left(\omega-\omega_c\right)(t-\tau)} c_\lambda^{\dagger}(\tau)=\kappa_\lambda \int_0^t d \tau \delta(t-\tau) c_\lambda^{\dagger}(\tau)=\frac{\kappa_\lambda}{2} c_\lambda^{\dagger}(t)
\end{equation}
\begin{equation}
\begin{gathered}
\frac{\sqrt{\kappa_\lambda \gamma_\lambda}}{2 \pi} \int_0^t d \tau \int d \omega e^{i\left(\omega-\omega_c\right)(t-\tau)} e^{-i \omega x_{j 0} / v} c_\lambda^{\dagger}(\tau)=\sqrt{\kappa_\lambda \gamma_\lambda} \int_0^t d \tau \delta\left(t-\frac{x_{j 0}}{v}-\tau\right) e^{-i \omega_c x_{j 0} / v} c_\lambda^{\dagger}(\tau) \\
\approx \sqrt{\kappa_\lambda \gamma_\lambda} \Theta\left(t-\frac{x_{j 0}}{v}\right) e^{-i k_\lambda x_{j 0}} c_\lambda^{\dagger}(t)
\end{gathered}
\end{equation}
\begin{equation}
\begin{gathered}\label{eqa11}
\frac{\gamma_\lambda}{2 \pi} \sum_{l=1}^N \int_0^t d \tau \int d \omega e^{i\left(\omega-\omega_c\right)(t-\tau)} e^{-i \omega x_{j l} / v} \sigma_{+}^{(l)}(\tau)=\gamma_\lambda \sum_{l=1}^N \int_0^t d \tau \delta\left(t-\frac{x_{j l}}{v}-\tau\right) e^{-i \omega_c x_{j l} / v} \sigma_{+}^{(l)}(\tau) \\
\approx \frac{\gamma_\lambda}{2} \sigma_{+}^{(j)}(t)+\gamma_\lambda \sum_{l=1}^N \Theta\left(t-\frac{x_{j l}}{v}\right) e^{-i k_\lambda x_{j l}} \sigma_{+}^{(l)}(t)
\end{gathered}
\end{equation}
where $x_{j0},x_{jl}>0$. $\Theta(t)$ is the step function. Substituting Eqs. (\ref{eqa9})-(\ref{eqa11}) into Eq. (\ref{eqa8}) and taking the averages, we obtain
\begin{equation}\label{eqa12}
\begin{aligned}
\frac{d}{d t}\langle O(t)\rangle= & \sum_{\lambda=R, L} \frac{\kappa_\lambda}{2}\left\{\left\langle c_\lambda^{\dagger}(t)\left[O(t), c_\lambda(t)\right]\right\rangle-\left\langle\left[O(t), c_\lambda^{\dagger}(t)\right] c_\lambda(t)\right\rangle\right\} \\
& +\sum_{\lambda=R, L} \sum_{j=1}^N \frac{\gamma_\lambda}{2}\left\{\left\langle\sigma_{+}^{(j)}(t)\left[O(t), \sigma_{-}^{(j)}(t)\right]\right\rangle-\left\langle\left[O(t), \sigma_{+}^{(j)}(t)\right] \sigma_{-}^{(j)}(t)\right\rangle\right\} \\
& +\sum_{\lambda=R, L} \sum_{j, l=1}^N \gamma_\lambda\left\{e^{-i k_\lambda x_{j l}}\left\langle\sigma_{+}^{(l)}(t)\left[O(t), \sigma_{-}^{(j)}(t)\right]\right\rangle-e^{i k_\lambda x_{j l}}\left\langle\left[O(t), \sigma_{+}^{(j)}(t)\right] \sigma_{-}^{(l)}(t)\right\rangle\right\} \\
& +\sum_{\lambda=R, L} \sum_{j=1}^N \sqrt{\kappa_\lambda \gamma_\lambda}\left\{e^{-i k_\lambda x_{j 0}}\left\langle c_\lambda^{\dagger}(t)\left[O(t), \sigma_{-}^{(j)}(t)\right]\right\rangle-e^{i k_\lambda x_{j 0}}\left\langle\left[O(t), \sigma_{+}^{(j)}(t)\right] c_\lambda(t)\right\rangle\right\}
\end{aligned}
\end{equation}
Since $\langle O(t)\rangle = \mathrm{Tr}\left[O(t)\rho(0)\right]=\mathrm{Tr}\left[O\rho(t)\right]$, we can simplify the average of operators in the above equation by using the cyclic property of trace. For example, the terms in the first and last lines of Eq. (\ref{eqa12}) can be written as
\begin{equation}
\left\langle c_\lambda^{\dagger}(t)\left[O(t), c_\lambda(t)\right]\right\rangle=\operatorname{Tr}\left[c_\lambda^{\dagger} O c_\lambda \rho(t)-c_\lambda^{\dagger} c_\lambda O \rho(t)\right]=\operatorname{Tr}\left[O c_\lambda \rho(t) c_\lambda^{\dagger}-O \rho(t) c_\lambda^{\dagger} c_\lambda\right]=\operatorname{Tr}\left\{O\left[c_\lambda, \rho(t) c_\lambda^{\dagger}\right]\right\}
\end{equation}
\begin{equation}
\left\langle\left[O(t), c_\lambda^{\dagger}(t)\right] c_\lambda(t)\right\rangle=\operatorname{Tr}\left[O c_\lambda^{\dagger} c_\lambda \rho(t)-c_\lambda^{\dagger} O c_\lambda \rho(t)\right]=\operatorname{Tr}\left[O c_\lambda^{\dagger} c_\lambda \rho(t)-O c_\lambda \rho(t) c_\lambda^{\dagger}\right]=\operatorname{Tr}\left\{O\left[c_\lambda^{\dagger}, c_\lambda \rho(t)\right]\right\}
\end{equation}
\begin{equation}
\begin{gathered}
\left\langle c_\lambda^{\dagger}(t)\left[O(t), \sigma_{-}^{(j)}(t)\right]\right\rangle=\operatorname{Tr}\left[c_\lambda^{\dagger} O \sigma_{-}^{(j)} \rho(t)-c_\lambda^{\dagger} \sigma_{-}^{(j)} O \rho(t)\right]=\operatorname{Tr}\left[O \sigma_{-}^{(j)} \rho(t) c_\lambda^{\dagger}-O \rho(t) c_\lambda^{\dagger} \sigma_{-}^{(j)}\right] \\
=\operatorname{Tr}\left\{O\left[\sigma_{-}^{(j)}, \rho(t) c_\lambda^{\dagger}\right]\right\}
\end{gathered}
\end{equation}
\begin{equation}
\begin{gathered}
\left\langle\left[O(t), \sigma_{+}^{(j)}(t)\right] c_\lambda(t)\right\rangle=\operatorname{Tr}\left[O \sigma_{+}^{(j)} c_\lambda \rho(t)-\sigma_{+}^{(j)} O c_\lambda \rho(t)\right]=\operatorname{Tr}\left[O \sigma_{+}^{(j)} c_\lambda \rho(t)-O c_\lambda \rho(t) \sigma_{+}^{(j)}\right] \\
=\operatorname{Tr}\left\{O\left[\sigma_{+}^{(j)}, c_\lambda \rho(t)\right]\right\}
\end{gathered}
\end{equation}
Therefore, we can obtain a quantum master equation in the following form 
\begin{equation}
\begin{aligned}
\frac{d}{d t} \rho(t)=-i[H, \rho(t)] & +\sum_{\lambda=R, L} \frac{\kappa_\lambda}{2}\left\{\left[c_\lambda, \rho(t) c_\lambda^{\dagger}\right]-\left[c_\lambda^{\dagger}, c_\lambda \rho(t)\right]\right\} \\
& +\sum_{\lambda=R, L} \sum_{j=1}^N \frac{\gamma_\lambda}{2}\left\{\left[\sigma_{-}^{(j)}, \rho(t) \sigma_{+}^{(j)}\right]-\left[\sigma_{+}^{(j)}, \sigma_{-}^{(j)} \rho(t)\right]\right\} \\
& +\sum_{\lambda=R, L} \sum_{j, l=1}^N \gamma_\lambda\left\{e^{-i k_\lambda x_{j l}}\left[\sigma_{-}^{(j)}, \rho(t) \sigma_{+}^{(l)}\right]-e^{i k_\lambda x_{j l}}\left[\sigma_{+}^{(j)}, \sigma_{-}^{(l)} \rho(t)\right]\right\} \\
& +\sum_{\lambda=R, L} \sum_{j, l=1}^N \sqrt{\kappa_\lambda \gamma_\lambda}\left\{e^{-i k_\lambda x_{j 0}}\left[\sigma_{-}^{(j)}, \rho(t) c_\lambda^{\dagger}\right]-e^{i k_\lambda x_{j 0}}\left[\sigma_{+}^{(j)}, c_\lambda \rho(t)\right]\right\}
\end{aligned}
\end{equation}
Note that we define $x_0=0$ in the main text, thus $x_{j0}=x_j$ in the last line on the right-hand side. In addition, the second and third terms on the right-hand side can be expressed using the Liouvillian superoperator. Taking into account the free-space decay $\gamma_0$ of atoms, we arrive at the extended cascaded quantum master equation given in Eqs. (\ref{eq1}) and (\ref{eq2}) in the main text.

\section{Inclusion of planewave excitation in the extended cascaded quantum master equation and the input-output boundary condition}\label{aaa}

The extended cascaded quantum master equation in Eqs. (\ref{eq1})-(\ref{eq6}) does not account for the excitation of the system. In this subsection, we consider a specific excitation configuration, i.e., planewave excitation through the right-propagating guided mode of waveguide ($b_R$), and derive the corresponding quantum master equation. To include the excitation, we start by formally integrating Eq. (\ref{eqa5}) and obtain
\begin{equation}\label{eqb1}
b_R(t)=b_R(0)+\int_0^t d \tau \sqrt{\frac{\kappa_R}{2 \pi}} c_R(\tau) e^{i\left(\omega-\omega_c\right) \tau} e^{-i \omega x_0 / v}+\sum_{j=1}^N \int_0^t d \tau \sqrt{\frac{\gamma_R}{2 \pi}} \sigma_{-}^{(j)}(\tau) e^{i\left(\omega-\omega_j\right) \tau} e^{-i \omega x_j / v}
\end{equation}
here $b_R (0)\neq0$ due to the existence of incident waveguide photons. Substituting the above equation into Eq. (\ref{eqa7}), we find the additional terms compared to Eq. (\ref{eqa8})
\begin{equation}
\sqrt{\frac{\kappa_R}{2 \pi}} \int d \omega\left\{b_R^{\dagger}(0) e^{i\left(\omega-\omega_c\right) t} e^{-i \omega x_0 / v}\left[O(t), c_R(t)\right]-\left[O(t), c_R^{\dagger}(t)\right] b_R(0) e^{-i\left(\omega-\omega_c\right) t} e^{i \omega x_0 / v}\right\}
\end{equation}
and
\begin{equation}
\sqrt{\frac{\gamma_R}{2 \pi}} \sum_{j=1}^N \int d \omega\left\{b_R^{\dagger}(0) e^{i\left(\omega-\omega_j\right) t} e^{-i \omega x_j / v}\left[O(t), \sigma_{-}^{(j)}(t)\right]-\left[O(t), \sigma_{+}^{(j)}(t)\right] b_R(0) e^{-i\left(\omega-\omega_j\right) t} e^{i \omega x_j / v}\right\}
\end{equation}
which yields the following Lindblad operator for excitation
\begin{equation}
\mathcal{D}_p[\rho]=\left[c_R, \rho_c(t)\right]-\left[c_R^{\dagger}, \rho_c^{\dagger}(t)\right]+\sum_{j=1}^N\left\{\left[\sigma_{-}^{(j)}, \rho_e(t)\right]-\left[\sigma_{+}^{(j)}, \rho_e^{\dagger}(t)\right]\right\}
\end{equation}
where $\rho_c (t)=\rho(t) p_c^{\dagger}$ and $\rho_e (t)=\rho(t) p_e^{\dagger}$ describe the driving from the incident source for CCW mode and atoms, respectively, with $p_c(t)=\sqrt{\frac{\kappa_R}{2 \pi}} \int d \omega b_R(0) e^{-i\left(\omega-\omega_c\right) t} e^{i \omega x_0 / v}$ and $p_e(t)=\sqrt{\frac{\gamma_R}{2 \pi}} \int d \omega b_R(0) e^{-i\left(\omega-\omega_j\right) t} e^{i \omega x_j / v}$ accounting for the absorption of the incident waveguide photons. We can see that for a monochromatic planewave, $p_c$ and $p_e$ reduce to a complex number. In this case, we have
\begin{equation}
\mathcal{D}_p[\rho]=\left[c_R, \rho(t)\right] p_c^*-\left[c_R^{\dagger}, \rho(t)\right] p_c+\sum_{j=1}^N\left\{\left[\sigma_{-}^{(j)}, \rho(t)\right] p_e^*-\left[\sigma_{+}^{(j)}, \rho(t)\right] p_e\right\}
\end{equation}
Accordingly, the extended cascaded quantum master equation is given by
\begin{equation}
\frac{d }{dt}{\rho}=-i[H, \rho]+\mathcal{D}[\rho]+\mathcal{D}_p[\rho]
\end{equation}
which yields Eqs. (\ref{eq14})-(\ref{eq16}) in the main text.

To derive the input-output relations, we integrate Eq. (\ref{eqa5}) from $t$ to $t_f$ (i.e., $t_f>t$) and obtain
\begin{equation}
b_R(t)=b_R(t_f)+\int_t^{t_f} d \tau \sqrt{\frac{\kappa_R}{2 \pi}} c_R(\tau) e^{i\left(\omega-\omega_c\right) \tau} e^{-i \omega x_0 / v}+\sum_{j=1}^N \int_t^{t_f} d \tau \sqrt{\frac{\gamma_R}{2 \pi}} \sigma_{-}^{(j)}(\tau) e^{i\left(\omega-\omega_j\right) \tau} e^{-i \omega x_j / v}
\end{equation}
By comparing with Eq. (\ref{eqb1}), we have
\begin{equation}\label{eqb8}
b_R(t)=b_R(0)+\int_0^{t_f} d \tau \sqrt{\frac{\kappa_R}{2 \pi}} c_R(\tau) e^{i\left(\omega-\omega_c\right) \tau} e^{-i \omega x_0 / v}+\sum_{j=1}^N \int_0^{t_f} d \tau \sqrt{\frac{\gamma_R}{2 \pi}} \sigma_{-}^{(j)}(\tau) e^{i\left(\omega-\omega_j\right) \tau} e^{-i \omega x_j / v}
\end{equation}
We use the following definition of input-output operators \cite{RN2,LZY}
\begin{equation}\label{eqb9}
b_{\text{out}}(t)=\frac{1}{\sqrt{2 \pi}} \int_0^{\infty} d \omega b_R\left(t_f\right) e^{i\left(\omega-\omega_c\right) x_N/v} e^{-i\left(\omega-\omega_c\right) t}
\end{equation}
\begin{equation}\label{eqb10}
b_{\text{in}}(t)=\frac{1}{\sqrt{2 \pi}} \int_0^{\infty} d \omega b_R\left(0\right) e^{-i\left(\omega-\omega_c\right) t}
\end{equation}
where a phase factor corresponding to the light propagating from the waveguide-cavity junction to the rightmost atom appears in Eq. (\ref{eqb9}). It means that the right output field propagates freely after scattered by the rightmost atom. Using Eqs. (\ref{eqb9}) and (\ref{eqb10}), we can obtain the input-output relation from Eq. (\ref{eqb8})
\begin{equation}
\begin{aligned}
\frac{1}{\sqrt{2 \pi}} \int_0^{\infty} d \omega b_R\left(t_f\right) & e^{i\left(\omega-\omega_c\right) x_N/v} e^{-i\left(\omega-\omega_c\right) t} \\
&=\frac{1}{\sqrt{2 \pi}} \int_0^{\infty} d \omega b_R(0) e^{i\left(\omega-\omega_c\right) x_N/v} e^{-i\left(\omega-\omega_c\right) t} \\
&+\frac{1}{\sqrt{2 \pi}} e^{i\left(\omega-\omega_c\right) x_N/v} \int_0^{t_f} d \tau \int_0^{\infty} d \omega \sqrt{\frac{\kappa_R}{2 \pi}} c_R(\tau) e^{-i\left(\omega-\omega_c\right)(t-\tau)} e^{-i \omega x_0/v} \\
&+\frac{1}{\sqrt{2 \pi}} e^{i\left(\omega-\omega_c\right) x_N/v} \sum_{j=1}^N \int_0^t d \tau \int_0^{\infty} d \omega \sqrt{\frac{\gamma_R}{2 \pi}} \sigma_{-}^{(j)}(\tau) e^{-i\left(\omega-\omega_c\right)(t-\tau)} e^{-i \omega x_j/v}
\end{aligned}
\end{equation}
\begin{equation}
\begin{aligned}
b_{\text{out}}(t)=b_{\text{in}} & \left(t-\frac{x_N}{v}\right)+\frac{1}{\sqrt{2 \pi}} \int_0^{t_f} d \tau \int_0^{\infty} d \omega \sqrt{\frac{\kappa_R}{2 \pi}} c_R(\tau) e^{-i\left(\omega-\omega_c\right)\left(t-\tau-x_{N 0} / v\right)} e^{-i \omega_c x_0/v} \\
& +\frac{1}{\sqrt{2 \pi}} \sum_{j=1}^N \int_0^{t_f} d \tau \int_0^{\infty} d \omega \sqrt{\frac{\gamma_R}{2 \pi}} \sigma_{-}^{(j)}(\tau) e^{-i\left(\omega-\omega_c\right)\left(t-\tau-x_{N j} / v\right)} e^{-i \omega_c x_j/v}
\end{aligned}
\end{equation}
\begin{equation}
\begin{aligned}
b_{\text{out}}(t)=b_{\text{in}} & \left(t-\frac{x_N}{v}\right)+\sqrt{\kappa_R} \int_0^{t f} d \tau c_R(\tau) \delta\left(t-x_{N 0} / v-\tau\right) e^{-i \omega_c x_0/v} \\
& +\sqrt{\gamma_R} \sum_{j=1}^N \int_0^{t_f} d \tau \sigma_{-}^{(j)}(\tau) \delta\left(t-x_{N j} / v-\tau\right) e^{-i \omega_c x_j/v}
\end{aligned}
\end{equation}
Applying the Markovian approximation, the above equation becomes
\begin{equation}\label{eqb14}
b_{\text{out}}(t) \approx b_{\text{in}}(t)+\sqrt{\kappa_R} c_R(t) e^{-i k_R x_0}+\sqrt{\gamma_R} \sum_{j=1}^N \sigma_{-}^{(j)}(t) e^{-i k_R x_j}
\end{equation}
With a fashion similar to Eqs. (\ref{eqb8})-(\ref{eqb14}), we can obtain the input-output relation for left-propagating guided mode
\begin{equation}\label{eqb15}
a_{\text{out}}(t) \approx a_{\text{in}}(t)+\sqrt{\kappa_L} c_L(t) e^{i k_L x_0}+\sqrt{\gamma_L} \sum_{j=1}^N \sigma_{-}^{(j)}(t) e^{i k_L x_j}
\end{equation}
with 
\begin{equation}
a_{\text{out}}(t)=\frac{1}{\sqrt{2 \pi}} \int_{-\infty}^0 d \omega b_L\left(t_f\right) e^{-i\left(\omega-\omega_c\right) x_N/v} e^{-i\left(\omega-\omega_c\right) t}
\end{equation}
and
\begin{equation}\label{eqb17}
a_{\text{in}}(t)=\frac{1}{\sqrt{2 \pi}} \int_{-\infty}^0 d \omega b_L\left(0\right) e^{-i\left(\omega-\omega_c\right) t}
\end{equation}
where a phase factor is added in $a_\text{out} \left(t\right)$ since the left output field propagates freely after scattered by the waveguide-cavity junction. Since $a_\text{in} \left(t\right)=0$ for left-incident planewave, the input-output relations Eqs. (\ref{eqb14}) and (\ref{eqb15}) yield Eqs. (\ref{eq18})-(\ref{eq21}) in the main text. Note that $a_\text{in} \left(t\right)$ [Eq. (\ref{eqb17})] is different from the field amplitude $a_\text{in}$ in Eq. (\ref{eq16}). 

\begin{figure}[t]
\centering\includegraphics[width=0.71\linewidth]{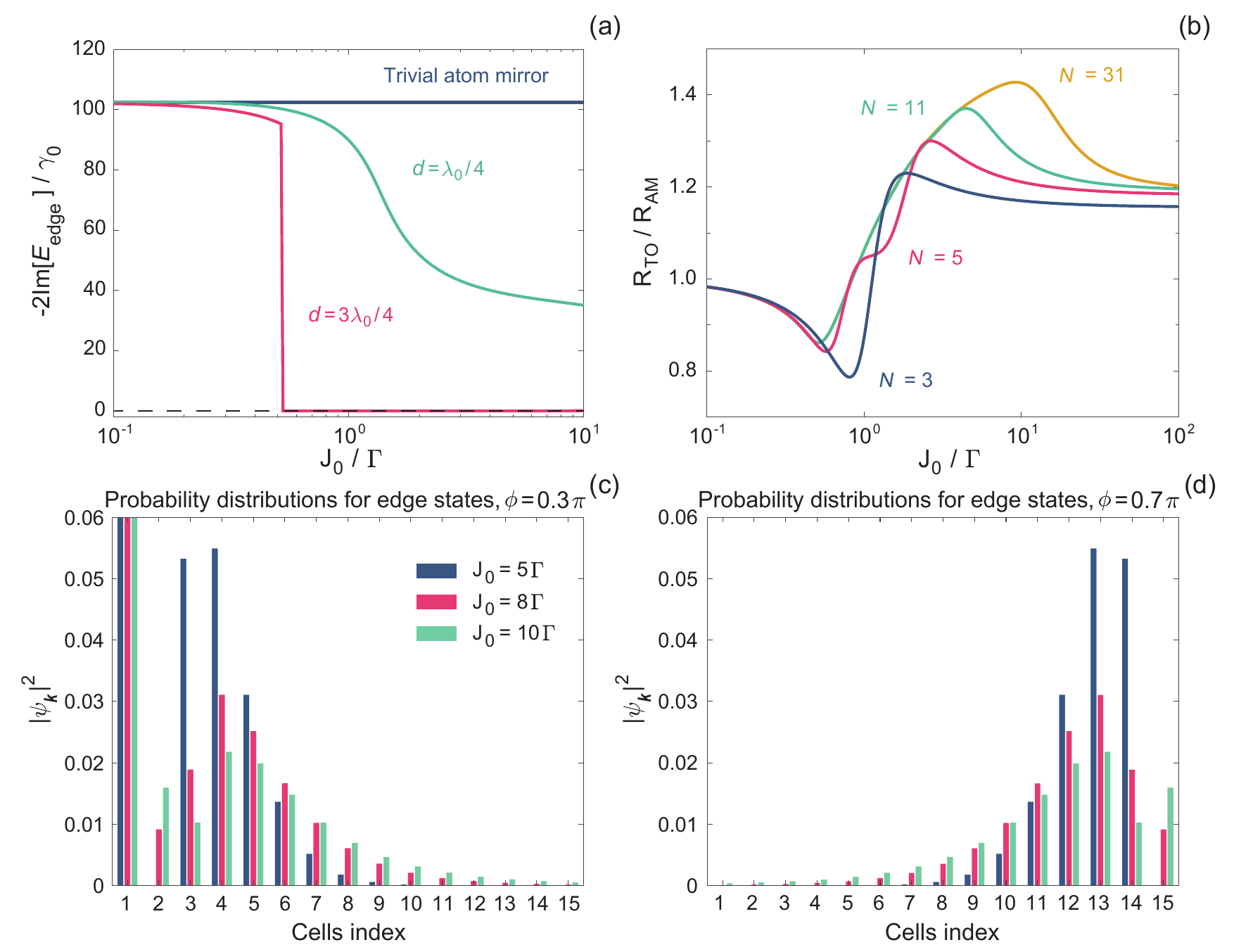}
\caption{(a) Decay rates of edge states with two atom spacings, $d=\lambda_0/4$ and $3\lambda_0/4$. The latter becomes dissipationless when $J_0>\Gamma/2$. The horizontal dashed black line indicates the zero decay rate. (b) Enhancement of reflection versus intensity strength $J_0$ for bare topological atom mirror with different number of atoms $N$. $R_\mathrm{TO}$ and $R_\mathrm{AM}$ are the reflection of topological and trivial atom mirrors, respectively. $R_\mathrm{AM}\approx0.68$. (c) and (d) Probability distributions of edge states for topological atom mirror with $\phi=0.3\pi$ and $0.7\pi$, respectively. Parameters not mentioned are the same as Fig. \ref{fig2} in the main text, while $\gamma_0=0$ in (a). }
\label{figs2}
\end{figure}

\section{Dissipative and dissipationless edge states of bare topological atom mirror}\label{ab}

To demonstrate the dissipationless feature of edge states, the SE rate of atoms is omitted, i.e., $\gamma_0=0$. In Fig. \ref{figs2}(a), we plot the decay rates $\Gamma_\mathrm{edge}=-2\operatorname{Im}\left[E_\mathrm{edge} \right]$ of edge states with atom spacings $d=\lambda_0/4$ and $3\lambda_0/4$ versus intensity strength $J_0$. It shows that $\Gamma_\mathrm{edge}$ of edge states with $d=3\lambda_0/4$ suddenly drops to zero at a critical $J_0 \approx \Gamma/2$ regardless of $\phi$, demonstrating the dissipationless feature; while the edge states in dissipative topological phase ($d=\lambda_0/4$) manifests the distinct feature of slowly decreased $\Gamma_\mathrm{edge}$ without transition as $J_0$ increases. As we see in Figs. \ref{figs2}(c) and (d), the delocalization populates all unit cells for a sufficient large intensity strength ($J_0>10\Gamma$) and as a consequence, $\Gamma_\mathrm{edge} \neq 0$ for dissipationless edge states.


In Figs. \ref{figs2}(c) and (d), we investigate the probability distributions of edge states for a bare topological atom mirror, i.e., without coupling to the cavity QED system. We can see that the topological edge states localize at the left (right) boundary when $0\leq\phi\leq\pi/2$ ($\pi/2<\phi\leq\pi$). The long-range hoppings between topological atoms induced by waveguide destroys the exponential localization of edge states: the probability decreases non-monotonically from the boundary and extends to the opposite boundary, exhibiting strong delocalization with a large intensity strength $J_0$. The delocalization means the increased dissipation of atom mirror and therefore, we can see in Fig. \ref{figs2}(b) that it requires more atoms to achieve high reflection for a large $J_0$. But on the other hand, a large $J_0$ can produce a wide topological bandgap, which is beneficial to suppress the dissipation from edge states to bulk states and achieve high reflection. The balance between the delocalization and the dissipation induced by bulk states yields an optimal $J_0$ for the maximal reflection of topological atom mirror with a fixed number of atoms. 

We stress that for $\gamma_0=0$, both topological and trivial atom mirrors have unity reflectivity, while the cavity polaritons of latter are dissipative. Therefore, the formation of bound cavity polaritons cannot understood as a result of classical destructive interference between the cavity field and the reflected field through atom mirror. 

\begin{figure}[t]
\centering\includegraphics[width=0.69\linewidth]{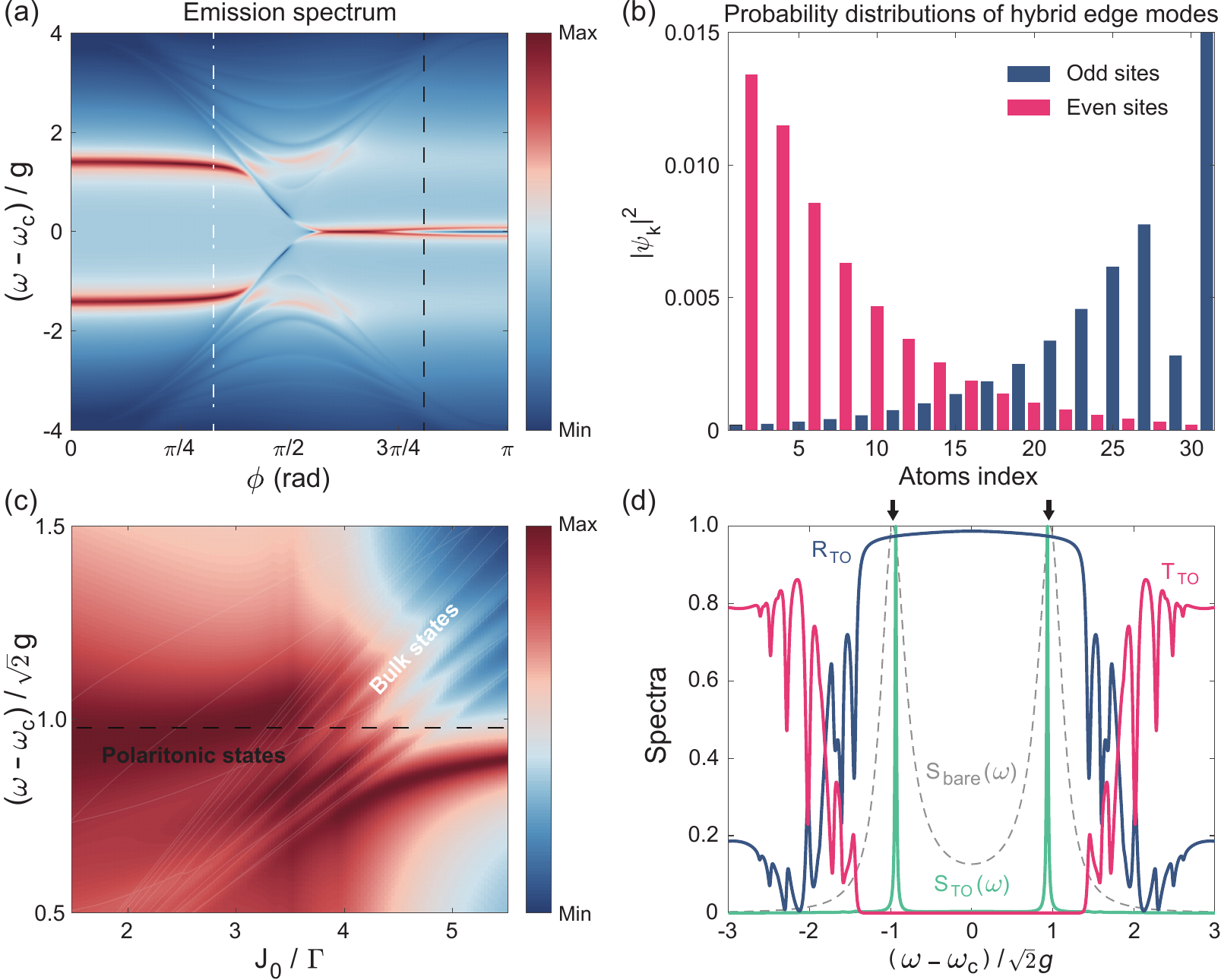}
\caption{(a) Emission spectrum of QE for cavity QED system with topological atom mirror versus $\phi$ under the strong-coupling regime. (b) Probability distributions of hybrid edge modes versus atoms index for $\phi=0.85\pi$ indicated by the dashed line in (a). Parameters are the same as Fig. \ref{fig3}(f). (c) A closeup of the emission spectrum of QE for the strong-coupling anticrossing between bulk states and polaritonic states, which are indicated by the black dashed line and the light gray lines, respectively. Parameters are the same as Fig. \ref{fig4}(b). (d) Comparison of the emission spectrum of QE and the reflection and transmission spectra for dark cavity polaritons in the planewave excitation, with $\phi$ indicated by the dashed dotted line in (a). }
\label{figs3}
\end{figure}

\section{Emission spectrum of QE and probability distributions of hybrid edge modes}\label{ad}

Fig. \ref{figs3}(a) presents the logarithmic plot of the emission spectrum of QE versus $\phi$. By comparing with the reflection spectrum shown in Fig. \ref{fig3}(f), we can see that they demonstrate similar pattern and features, such as the variation of topological bandgap and the distinct linewidth of cavity polaritons for the left and right edge states. However, for observing the linewidth narrowing of polaritonic states (Rabi peaks), the emission spectrum of QE is preferred since the signal of bulk states is weak. In addition, we find that the composed system shows a feature of SSH model with even sites, where two hybrid edge modes with a finite gap can be seen in the region of $\pi/2<\phi\leq\pi$, which is hard to recognize in the reflection spectrum. The probability distributions shown in Fig. \ref{figs3}(b) indicate the formation of hybrid edge modes with even and odd parities, distinguishing from the left or right edge states of bare topological atom mirror that populate either the odd or even sites.

On the other hand, we find an inconspicuous anticrossing at $\phi\sim0.4\pi$ in Fig. \ref{figs3}(a), indicating the strong coupling between bulk states and polaritonic states. This anticrossing behaviour is more evident in the emission spectrum of QE versus $J_0$, as Fig. \ref{figs3}(c) shows. In stark contrast to the conventional strong-coupling anticrossing that is observed by tuning the frequency detuning, here $J_0$ plays the role of frequency detuning in the strong-coupling anticrossing. It is because the variation of $J_0$ does not alter the energy of polaritonic states, but the width of topological bandgap linearly enlarges with increased $J_0$. Therefore, $J_0$ can change the energies of bulk states.

It is important to note that in the reflection spectrum of Fig. \ref{fig3}(f), two dips corresponding to cavity polaritons are vanishing around $\phi=0.33\pi$, while this phenomenon is not found in the emission spectrum of QE, as Fig. \ref{figs3}(d) shows, where the black arrows indicate the polaritonic states. The results reveal that the formation of dark cavity polaritons is related to specific excitation method. For the configuration of planewave excitation, these dark cavity polaritons stem from destructive interference between the incident and scattering photons.

\section{Dissipation spectrum and radiating modes}\label{ae}

With the matrix elements in the Lindblad operator Eq. (\ref{eq2}), we can obtain a dissipation matrix $\gamma$, which can be expressed as follows
\begin{equation}
\gamma=\sum_m \chi_m\left|v_m\right\rangle\left\langle v_m\right|
\end{equation}
where $\chi_m$ is called the dissipation spectrum and $\left|v_m\right\rangle$ is the corresponding wave function. Fig. \ref{figs4}(a) shows $\chi_m$ for the composed system in the strong-coupling regime with $31$ atoms in mirror. We can see that the dissipation of modes $m=1$ and $34$ is $\chi_{1,34}\sim\kappa/2$, thus they are related to two cavity modes. Two radiating modes indexed by $m=2$ and $3$ can be found in the dissipation spectrum, whose dissipation is much greater than other modes. Fig. \ref{figs4}(b) plots the wave function of two radiating modes versus atoms index, where we can identify the odd and even polarization for $m=2$ and $3$, respectively.  
With the eigenstates $\left|\psi_n\right\rangle$ of Hamiltonian $\operatorname{Re}\left[H_\mathrm{eff} \right]$, we can evaluate the dissipation rate $\Gamma_n$ of the $n$th eigenstate to the environment
\begin{equation}
\Gamma_n=\left\langle\psi_n|\gamma| \psi_n\right\rangle=\sum_m \Gamma_n^m
\end{equation}
with $\Gamma_n^m$ being the contribution of the $m$th mode in dissipation spectrum
\begin{equation}
\Gamma_n^m=\chi_m\left\langle\psi_n \mid v_m\right\rangle^2
\end{equation}
Particularly, the eigenstates corresponding to cavity polaritons is indicated by $n= \pm$. Fig. \ref{fig4}(d) shows the contributions of cavity modes ($\Gamma_\pm^{1}$ and $\Gamma_\pm^{34}$) and radiating modes ($\Gamma_\pm^{2}$ and $\Gamma_\pm^{3}$) to the dissipation of cavity polaritons. 

\begin{figure}[H]
\centering\includegraphics[width=0.7\linewidth]{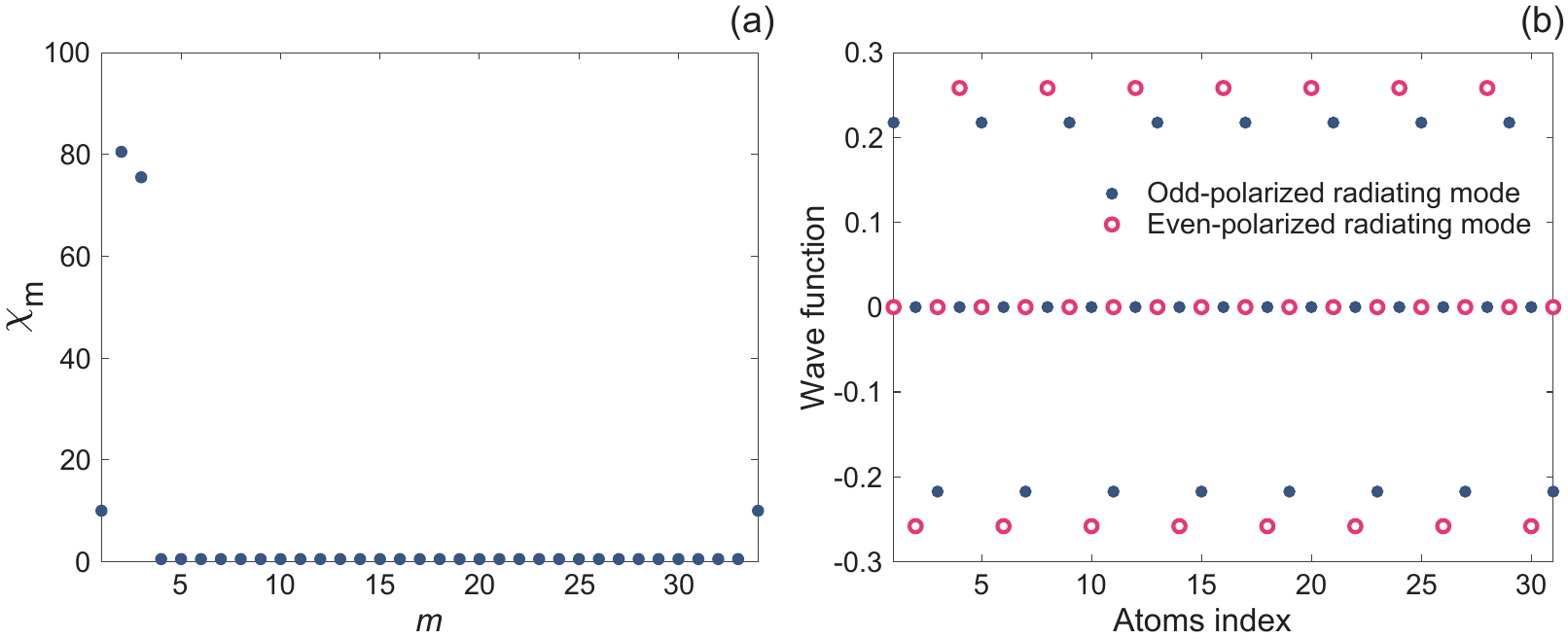}
\caption{Dissipation spectrum (a) and the wave function of two radiating modes (b) of topological atom mirror. The parameters are the same as Fig. \ref{fig4}(d).  }
\label{figs4}
\end{figure}
\end{widetext}

\nocite{*}

\bibliography{TO}

\end{document}